\documentclass[conference]{IEEEtran}
\IEEEoverridecommandlockouts
\usepackage{booktabs}   %% For formal tables:
\usepackage{subcaption} %% For complex figures with subfigures/subcaptions
\usepackage{array}
\usepackage{amsmath,amssymb,amsfonts}
\usepackage{algorithm}
\usepackage[noend]{algpseudocode}
\usepackage{graphicx}
\usepackage{textcomp}
\usepackage{float} 
\usepackage{listings}
\usepackage{xspace}
\usepackage{amsthm}
\usepackage{multirow}
\usepackage{url}
\usepackage{balance}

\usepackage{enumitem}
\usepackage[skins]{tcolorbox}

\newtcolorbox{myframe}[2][]{%
  enhanced,colback=white,colframe=black,coltitle=black,
  sharp corners,
  toprule=1.0pt,
  rightrule=0.3pt,
  leftrule=0pt,
  bottomrule=0pt,
  fonttitle=\itshape\scshape\large,
  left=0pt,right=5pt,top=5pt,bottom=3pt,
  attach boxed title to top right={yshift=-0.3\baselineskip-0.4pt,xshift=-5mm},
  boxed title style={tile,size=minimal,left=0.2mm,right=0.5mm,
    colback=white,before upper=\strut},
  title=#2,#1
}

\newcommand{\tool}{\textsc{DeepRL4FL}\xspace}

\newcolumntype{L}[1]{>{\raggedright\arraybackslash}p{#1}}

\newcommand{\code}[1]{{\small\textsf{#1}}}
\usepackage{amsthm}
 \definecolor{dkgreen}{rgb}{0,0.6,0}
\definecolor{gray}{rgb}{0.5,0.5,0.5}
\definecolor{mauve}{rgb}{0.58,0,0.82}
\begin{document}

\title{Fault Localization with Code Coverage Representation Learning}

\author{\IEEEauthorblockN{Yi Li}
\IEEEauthorblockA{\textit{Department of Informatics} \\
  \textit{New Jersey Institute of Technology}\\
  New Jersey, USA \\
Email: yl622@njit.edu}
\and
\IEEEauthorblockN{Shaohua Wang\IEEEauthorrefmark{1}}
\IEEEauthorblockA{\textit{Department of Informatics} \\
  \textit{New Jersey Institute of Technology}\\
  New Jersey, USA \\
Email: davidsw@njit.edu 
}
\and
\IEEEauthorblockN{Tien N. Nguyen}
\IEEEauthorblockA{\textit{Computer Science Department} \\
  \textit{The University of Texas at Dallas}\\
  Texas, USA \\
Email: tien.n.nguyen@utdallas.edu}
}

\maketitle
\begingroup\renewcommand\thefootnote{\IEEEauthorrefmark{1}}
\footnotetext{Corresponding Author}
\endgroup
\begin{abstract}
In this paper, we propose {\tool}, a deep learning fault localization
(FL) approach that locates the buggy code at the statement and method
levels by treating FL as an image pattern recognition problem. {\tool}
does so via novel code coverage representation learning (RL) and data
dependencies RL for program statements. Those two types of RL on the
dynamic information in a code coverage matrix are also combined with
the code representation learning on the static information of the
usual suspicious source code. This combination is inspired by crime
scene investigation in which investigators analyze the crime scene
(failed test cases and statements) and related persons (statements
with dependencies), and at the same time, examine the usual suspects
who have committed a similar crime in the past (similar buggy code in
the training data).

%For code coverage representation learning,

For the code coverage information, {\tool} first orders the test cases
and marks error-exhibiting code statements, expecting that a model can
recognize the patterns discriminating between faulty and non-faulty
statements/methods.
For dependencies among statements, the suspiciousness of a statement
is seen taking into account the data dependencies to other statements
in execution and data flows, in addition to the statement by
itself.
Finally, the vector representations for code coverage matrix, data
dependencies among statements, and source code are combined and used
as the input of a classifier built from a Convolution Neural Network
to detect buggy statements/methods.
%{\tool} first enhance a code coverage matrix by ordering test cases
%using their relations and marking error-exhibiting code statements
%using the information from the failing test cases. Second, {\tool}
%dynamically and statically builds statement dependencies through
%execution paths and data-flow graphs. Third, {\tool} combines
%statement dependencies with the enhanced matrix for better fault
%localization.
%
Our empirical evaluation shows that {\tool} improves the top-1 results
over the state-of-the-art statement-level FL baselines from 173.1\% to
491.7\%. It also improves the top-1 results over the existing
method-level FL baselines from 15.0\% to 206.3\%. 

%Our empirical evaluation shows that \tool outperforms the baseline
%models and localizes 245 bugs from Defects4J. It improves the top-1
%results of baselines from 15.0\%--206.3\%.
\end{abstract}


\section{Introduction}

Finding and fixing software defects is an important process to ensure
a high-quality software product.
%
%Much time and effort from developers have been spent in that
%process.
To reduce developers' effort, several {\em fault localization} (FL)
approaches~\cite{fl-survey} have been proposed to help
localize the source of a defect (also called a {\em bug} or {\em
  fault}).  In the FL problem, given the execution of test cases, an
FL tool identifies the set of {\em suspicious lines of code} with
their associated suspiciousness scores~\cite{fl-survey}.  The key
input of an FL tool is the {\em code coverage matrix} in which the
rows and columns correspond to the source code statements and test
cases, respectively.  Each cell is assigned with the value of 1 if the
respective statement is executed in the respective test case, and with
the value of 0, otherwise. In recent FL, several researchers also
advocate for fault localization at method level~\cite{DeepFL}. FL at
both levels are useful for developers.
%in locating and fixing bugs.

%Traditionally,

{\em Spectrum-based fault localization} (SBFL)
approaches~\cite{Ochiai,jones2001visualization,keller2017critical}
take the recorded lines of code that were covered by each of the given
test cases, and assigned each line of code a suspiciousness score
based on the code coverage matrix. Despite using different
formulas to compute that score, the idea is that a line covered more
in the failing test cases than in the passing ones is more suspicious
than a line executed more in the passing ones. A key drawback of those
approaches is that the same score is given to the lines that have been
executed in both failing and passing test cases. An example is the
statements that are part of a block statement and executed at the
same nested level. Another example is the conditions of the
condition statements, e.g., \code{if}, \code{while}, \code{do},
and \code{switch}. 

To improve SBFL, {\em mutation-based fault localization} (MBFL)
approaches~\cite{MUSE,papadakis2012using,Metallaxis}
enhance the code coverage information by modifying a statement with
mutation operators, and then collecting code coverages when executing
the mutated programs with the test cases. They apply suspiciousness
score formulas in the same manner as the spectrum-based FL approaches on
the code coverage matrix for each original statement and its mutated
ones. Despite the improvement, MBFL are not effective for the bugs
that require the fixes that are more complex than a mutation
(Section~\ref{motivexample}).

{\em Machine learning (ML)} and {\em deep learning (DL)}
have been used in fault localization. DeepFL~\cite{DeepFL}
computes for each faulty method a vector with +200 scores in which
each score is computed via a specific feature, e.g., a spectrum-based
or mutation-based formula, or a code complexity metric. Despite its
success, the accuracy of DeepFL is still limited. A reason could be
that it uses various calculated scores from different formulas as a
proxy to learn the suspiciousness of a faulty element, instead of
fully exploiting the code coverage. Some formulas, such
as the spectrum- and mutation-based formulas, inherently suffer from
the issues as explained earlier with the statements covered by
both failing and passing test cases.

%, even they are mixed with other types of formulas.
%Another reason is that the scores computed by different formula are
%independent. Putting the columns corresponding to those 34 scores in a
%different order might have different results.

%In this paper,

We propose {\tool}, a fault localization approach for buggy
statements/methods that exploits the image classification and pattern
recognition capability of the Convolution Neural Network
(CNN)~\cite{krizhevsky-2012} to apply on the code coverage (CC)
matrix. Instead of summarizing each row in that matrix with a
suspiciousness score, we use its full details. Importantly, we enhance
the matrix to facilitate the application of the CNN model in {\em
  recognizing the key characteristics in the matrix} to discriminate
more easily between faulty and non-faulty statements/methods. Toward
that end, we order the columns (test cases) of the CC matrix so that
the {\em test cases with the non-zero values on nearby statements are
  close together}.
Specifically, the first test case covers the most statements. The
next test case shares with the previous one as many executed
statements as possible.
%
%This puts {\em the non-zero cells in the matrix close together}.
%
We expect that {\em the CNN model with its capability to learn the
  relationships among nearby cells via a small filter} can recognize
the visual characteristic features to discriminate faulty and
non-faulty statements/methods.

Inspired by the method in crime scene investigation, we~use three
sources of information for FL: 1) code coverage matrix with failed
test cases (the crime scene and victims), 2) similar buggy code in the
history ({\em usual suspects} who have committed a similar crime in
the past), and 3) the statements with data dependencies (related
persons). First, the evidences at the crime scene are always examined.
For an analogy, the CC matrix for the occurrence of the fault is
analyzed.
%the CC matrix for the occurrence of the fault is an analogy of the
%evidences at the scene.
%
Second, an investigator also makes a connection from the crime scene
to {\em the usual suspects}.  This is analogous~to the modeling of the
code of the faults that have been encountered in the training
dataset. The idea is that if the persons (analogous to the code) who
have committed the crimes with similar modus operandi (M.O.)  in the
past are likely the suspects (code with high suspiciousness) in the
current investigation.

Third, in addition to the crime scene, the investigator also looks at
the relationships between the victim or the things~happening at the
scene and other related persons. Thus, in addition to the statement
itself, its suspiciousness is viewed taking into
account the data dependencies to other statements in execution flows
and data flows. The idea is that some statements, even far away from
the buggy line, could have impacts or exhibit the consequences of the
buggy line when they are data-dependent.
%Those lines are more helpful in locating the buggy line than the
%nearby lines but without data dependencies.
Thus, for a test, we first identify the error-exhibiting (EE) line
(defined as the line where the program crashed or exhibited an
incorrect value(s)/behavior(s)). That is, if the program crashes, the
error-exhibiting line is listed. If there is no crash and an assertion
fails, assertion statement is EE line. EE line is usually specified in
a test execution. To identify the related statements, from the EE
line, we consider the execution order. However, if the statements are
in the same block of code (i.e., being executed sequentially), we also
consider the data dependencies among them and with the EE line.
%----------------------------------
Finally, all three sources of information are encoded into
vector/matrix representations, which are used as input to the CNN
model to act as a classifier to decide whether a statement/method
as a faulty or not.

%Tien removed
%Specifically, the code coverage matrix is enhanced and encoded into
%the spectrum-based and/or mutation-based vector representations. To
%model the data dependencies and execution orders among statements, we
%use word2vec~\cite{word2vec} and node2vec~\cite{Grover-2016}. To
%capture the usual suspicious code, we encode the AST via the paths
%across the subtrees to represent the source code.

%To feed as the input to the CNN model for buggy or not buggy
%classification, we use three types of representations for different
%information. First, we use the dynamic information from the test
%execution by encoding the enhanced code coverage matrix into
%spectrum-based and/or mutation-based vector representations. The
%second input is from the combination of dynamic and static
%information, in which we encode the data dependencies and the
%execution order among the statements via word2vec~\cite{word2vec} and
%node2vec~\cite{Grover-2016}. The final input is from the static
%analysis on the abstract syntax tree (AST). Specifically, we encode
%the AST via the paths across the subtrees to represent the source
%code. Finally, all the representation vectors including spectrum-based
%representation vectors and/or mutation-based representation vectors,
%and the statement representation vectors are used as the inputs of a
%CNN model that acts as the classifier for each statement or each
%method as buggy or not.

We conducted several experiments to evaluate {\tool} on Defects4J
benchmark~\cite{defects4j}. Our empirical results show that \tool
locates 245 faults and 71 faults at the method level and the statement
level, respectively, using only top-1 candidate (i.e., the first
ranked element is faulty). It can improve the top-1 results of the
state-of-the-art \textit{statement-level} FL baselines by 317.7\%,
273.7\%, 173.1\%, 195.8\%, and 491.7\% when comparing with
Ochiai~\cite{Ochiai}, Dstar~\cite{DStar}, Muse~\cite{MUSE},
Metallaxis~\cite{Metallaxis}, and RBF-Neural-Network-based FL
(RBF)~\cite{RBF_Neural_Network}, respectively.  {\tool} also improves
the top-1 results of the existing \textit{method-level} FL baselines,
MULTRIC~\cite{MULTRIC}, FLUCCS~\cite{FLUCCS}, TraPT~\cite{TraPT}, and
DeepFL~\cite{DeepFL}, by 206.3\%, 53.1\%, 57.1\%, and 15.0\%,
respectively. Our results show that three sources of information in
{\tool} positively contribute to its high accuracy.

%Our sensitivity analysis shows that the three above types of
%information as parts of {\tool} contribute positively toward its
%higher accuracy.

We also evaluated {\tool} on ManyBugs~\cite{LeGoues15tse}, a~ben\-chmark
of C code with 9 projects. The results are~consistent with the ones
on Java code.  {\tool} localizes 27 faulty statements and 98 faulty
methods using only top-1 results.

%the components designed to implement our key ideas can also contribute
%to our \tool.

The contributions of this paper are listed as follows:
%\begin{itemize}
	%\vspace{5pt}

{\bf 1. Novel code coverage representation.} Our representation
enables fully exploiting test coverage matrix and taking advantage of
the CNN model in image recognition to localize~faults.

%by analyzing the relations among test cases, among code statements,
%among test cases and code statements.

{\bf 2. {\tool}: Novel DL-based fault localization approach.} Test
case ordering and three sources of information allow treating FL as a
pattern recognition. Without ordering and statement dependencies, the
CNN model will not work~well.

%Inspired by crime scene investigation, we integrate into {\tool} the
%information from the code coverage matrix, the inter-related
%statements via data dependencies and execution order, as well as the
%information from usual suspicious source code.

%\tool learns to represent code coverage information, statements
%dependencies, and source code, and combines them into representative
%vectors for CNN-based fault localization.

{\bf 3. Extensive empirical evaluation.} We evaluated our model against the most
recent FL models at the statement and method levels, in both
within-project and cross-project settings, and for both C and Java.
%{\color{blue}{Furthermore, {\tool} has been tested on Java and C projects.}}
%Our \tool outperforms all studied baselines at both statement and
%method levels and under two scenarios.
Our replication package is available at~\cite{FaultLocalization2021}.

%\end{itemize}

%\vspace{-0.05in}
\section{Motivating Examples}
\label{motivexample}

%\subsection{Motivating Examples}\label{motivexample}

\begin{figure}[t]
%	\vspace{-10pt}
	\centering
	\lstset{
		numbers=left,
		numberstyle= \tiny,
		keywordstyle= \color{blue!70},
		commentstyle= \color{red!50!green!50!blue!50},
		frame=shadowbox,
		rulesepcolor= \color{red!20!green!20!blue!20} ,
		xleftmargin=1.5em,xrightmargin=0em, aboveskip=1em,
		framexleftmargin=1.5em,
                numbersep= 5pt,
		language=Java,
%		basicstyle=\footnotesize, %\scriptsize\ttfamily,
%     basicstyle=\scriptsize\sffamily,
    basicstyle=\scriptsize\ttfamily,
    numberstyle=\scriptsize\ttfamily,
    emphstyle=\bfseries,
                moredelim=**[is][\color{red}]{@}{@},
		escapeinside= {(*@}{@*)}
	}
	\begin{lstlisting}[]
public static String join(Object[] array, char separator, 
   int startIndex, int endIndex) {
  if (array == null) {
    return null;
  }
  int noOfItems = (endIndex - startIndex);
  if (noOfItems <= 0) {
     return EMPTY;
  }
(*@{\color{red}{-StringBuilder buf = new StringBuilder((array[startIndex]}@*)
(*@{\color{red}{ == null? 16 : array[startIndex].toString().length())+1);}@*)
(*@{\color{cyan}{+ StringBuilder buf = new StringBuilder(noOfItems * 16);}@*)
  for (int i = startIndex; i < endIndex; i++) {
     if (i > startIndex) {
        buf.append(separator);
     }
     if (array[i] != null) {
        buf.append(array[i]);
     }
  }
  return buf.toString();
}	
	\end{lstlisting}
        \vskip -7pt
	\caption{An Example of a Buggy Statement}
        %Motivating Example 1}
\label{Fig:motiv_example_I}	
\vspace{-10pt}
\end{figure}

%Let us start with the examples motivating {\tool}. 
Fig.~\ref{Fig:motiv_example_I} shows a real-world example of a bug
in Defects4J~\cite{defects4j}.  The bug occurs at line 10 in which the
length of the string to be built via
\textit{StringBuilder} was not set correctly. 
A developer fixed the bug by modifying lines 10--11 into line~12.

To localize the buggy line, there exist three categories of
approaches. The first one is spectrum-based fault localization
(SBFL). The key idea in SBFL is that in a test dataset, {\em a line
executed more in the failing test cases than in the passing~ones is
considered as more suspicious than a line executed more~in the passing
ones}. A summary of the CC matrix for this bug~is shown in
Fig.~\ref{fig:motiv_example_I_b}. The lines 3, 6--7, and 10--11 in
Fig.~\ref{Fig:motiv_example_I}~are
executed in both passing and failing test cases, and as~a result,
given {\em the same suspiciousness scores}. Thus, SBFL is ineffective
to detect the buggy line 10 and this buggy~method.

%(*@{\color{red}{\quad\quad\quad\quad + 1);}@*)

\begin{figure}[t]
	\scriptsize
\centering
\renewcommand{\arraystretch}{0.8}
\tabcolsep 3pt
	\begin{subfigure}{0.26\textwidth}
	\centering
	\begin{tabular}{p{1.2cm}<{\centering}|p{0.05cm}<{\centering}p{0.05cm}<{\centering}p{0.05cm}<{\centering}p{0.05cm}<{\centering}p{0.05cm}<{\centering}p{0.05cm}<{\centering}p{0.05cm}<{\centering}p{0.05cm}<{\centering}p{0.05cm}<{\centering}p{0.05cm}<{\centering}p{0.05cm}<{\centering}}
		
		\hline
                & $t_{1}$ & $t_{2}$ &  &  &  & ... &  \\
		line-3 &$\circ$ &... & $\bullet$ & ... & $\bullet$ & ...& $\circ$ & ...&$\bullet$ &... & $\bullet$ \\
		line-4 &$\circ$ &... & $\circ$ & ... & $\circ$ & ...& $\circ$ & ...&$\circ$ &... & $\circ$ \\
		line-6 &$\circ$ &... & $\bullet$ & ... & $\bullet$ & ...& $\circ$ & ...&$\bullet$ &... & $\bullet$  \\
		line-7 &$\circ$ &... & $\bullet$ & ... & $\bullet$ & ...& $\circ$ & ...&$\bullet$ &... & $\bullet$  \\
		line-8 &$\circ$ &... & $\circ$ & ... & $\bullet$ & ...& $\circ$ & ...&$\circ$ &... & $\bullet$  \\
		line-(10-11) &$\circ$ &... & $\bullet$ & ... & $\bullet$ & ...& $\circ$ & ...&$\bullet$ &... & $\bullet$  \\
		line-13 &$\circ$ &... & $\bullet$ & ... & $\bullet$ & ...& $\circ$ & ...&$\circ$ &... & $\bullet$ \\
		line-14 &$\circ$ &... & $\bullet$ & ... & $\bullet$ & ...& $\circ$ & ...&$\circ$ &... & $\bullet$ \\
		line-15 &$\circ$ &... & $\bullet$ & ... & $\bullet$ & ...& $\circ$ & ...&$\circ$ &... & $\bullet$ \\
		line-17 &$\circ$ &... & $\bullet$ & ... & $\bullet$ & ...& $\circ$ & ...&$\circ$ &... & $\bullet$  \\
		line-18 &$\circ$ &... & $\bullet$ & ... & $\bullet$ & ...& $\circ$ & ...&$\circ$ &... & $\bullet$  \\
		line-21 &$\circ$ &... & $\bullet$ & ... & $\bullet$ & ...& $\circ$ & ...&$\circ$ &... & $\bullet$  \\
		\hline
	\end{tabular}
	\caption{}
	\label{fig:motiv_example_I_b}
	\end{subfigure}
	\begin{subfigure}{0.21\textwidth}
		\centering
		
	\begin{tabular}{p{1.2cm}<{\centering}|p{0.05cm}<{\centering}p{0.05cm}<{\centering}p{0.05cm}<{\centering}p{0.05cm}<{\centering}p{0.05cm}<{\centering}p{0.05cm}<{\centering}p{0.05cm}<{\centering}}
		
		\hline
                & $t_{9}$ & $t_{33}$ &  &  &  & ... &  \\
		line-3 &$\bullet$ & $\bullet$ & $\bullet$ & $\bullet$ & $\circ$ & ... & $\circ$ \\
		line-4 &$\circ$ & $\circ$ & $\circ$ & $\circ$ & $\circ$ & ... & $\circ$ \\
		line-6 &$\bullet$ & $\bullet$ & $\bullet$ & $\bullet$ & $\circ$ & ... & $\circ$  \\
		line-7 &$\bullet$ & $\bullet$ & $\bullet$ & $\bullet$ & $\circ$ & ... & $\circ$  \\
		line-8 &$\circ$ & $\circ$ & $\bullet$ & $\bullet$ & $\circ$ & ... & $\circ$  \\
		line-(10-11) &$\bullet$ & $\star$ & $\bullet$ & $\bullet$ & $\circ$ & ... & $\circ$  \\
		line-13 &$\star$ & $\bullet$ & $\bullet$ & $\bullet$ & $\circ$ & ... & $\circ$ \\
		line-14 &$\bullet$ & $\circ$ & $\bullet$ & $\bullet$ & $\circ$ & ... & $\circ$ \\
		line-15 &$\bullet$ & $\circ$ & $\bullet$ & $\bullet$ & $\circ$ & ... & $\circ$ \\
		line-17 &$\bullet$ & $\circ$ & $\bullet$ & $\bullet$ & $\circ$ & ... & $\circ$  \\
		line-18 &$\bullet$ & $\circ$ & $\bullet$ & $\bullet$ & $\circ$ & ... & $\circ$  \\
		line-21 &$\bullet$ & $\circ$ & $\bullet$ & $\bullet$ & $\circ$ & ... & $\circ$  \\
		\hline
	\end{tabular}
	\caption{}
	\label{fig:motiv_example_I_c}
	\end{subfigure}
\vspace{-8pt}
\renewcommand{\arraystretch}{1}
\captionsetup{justification=centering}
	\caption{Code Coverage for Fig.~\ref{Fig:motiv_example_I} (Note: $\circ$, $\bullet$, $\star$ for 0,1,-1)}
	\label{fig:motiv_example_cc}
	\vspace{-11pt}
\end{figure}

The second category is mutation-based fault localization (MBFL). A
MBFL approach (e.g., Metallaxis~\cite{Metallaxis}) modifies a
statement using mutation operators. After collecting code coverage
information for each statement regarding to multiple mutations, it
computes the suspiciousness score for each~statement using a
spectrum-based formula (e.g., Ochiai~\cite{Ochiai}) on the CC matrix
for each original statement and for its mutated ones. However, the
fix for the buggy line 10 requires more complex code transformations
than a mutation. Thus, an MBFL approach cannot detect
this buggy line and buggy~method.

The third category is deep learning and machine learning-based FL
approaches~\cite{DeepFL,RBF_Neural_Network}. Specifically, Wong {\em
el al.}~\cite{RBF_Neural_Network} use a backpropagation neural network
on code coverage for each statement. Since the lines 3, 6--7, and
10--11 are executed in both passing and failing test cases, the model
cannot learn to distinguish them to detect the buggy
line~10. DeepFL~\cite{DeepFL}, uses multilayer perceptron (MLP) on a
matrix in which each row corresponds to a statement, while each column
is a suspiciousness score computed by a spectrum-based formula, or a
code complexity metric.
%
%{\color{blue}{In total, they computed 200+ suspiciousness scores for
%an faulty element~\cite{li2019deepfl}.}}
%In total, they use 34 columns corresponding to 34 different
%spectrum-based formulas~\cite{li2019deepfl}. 
In our experiment (Section~\ref{rq1-results}), DeepFL could not detect
the buggy line 10.
Despite combining several scores, the aforementioned lines are
given the same suspiciousness scores by each spectrum-based
formula.
%Second, the formulas are {\em independent} of one another, thus,
%putting the scores from all formulas side by side to form a matrix
%creates randomness, making it harder for the MLP model to learn for
%FL.

\vspace{0.03in}
\noindent {\underline{{\bf Observation 1.}} The state-of-the-art
spectrum-based~\cite{keller2017critical,lucia2014extended},
mutation-based~\cite{MUSE,papadakis2012using, Metallaxis}, and deep
learning-based FL approaches~\cite{DeepFL} do not consider the full
details of the CC matrix. Instead, they summarize each statement/row
with a suspiciousness score, thus limiting their capabilities.

To address that, we aim to exploit the full details of the CC matrix
via the use of the CNN model~\cite{krizhevsky-2012}, which has been
shown to be effective in image pattern recognition. However, there is
a challenge: if we do not enforce an order on the test cases
(columns), we might end up with a CC matrix with the dark cells (the
values of 1) that are far apart
(Fig.~\ref{fig:motiv_example_I_b}). Note that {\em the CNN model is
effective to learn the relationships among the nearby cells in a
matrix with its small~sliding window
(called filter)}~\cite{krizhevsky-2012}. Thus, {\em we need to enforce an
order on the test cases, i.e., the columns of the CC matrix so that
the values of 1 on the same or nearby rows get to be close to one
another}. For example, if we enforce an order with the mentioned
strategy (we will explain the detailed algorithm later) for the
running example, we will have the matrix in
Fig.~\ref{fig:motiv_example_I_c}. That is, the results for the test
cases 9, 33, etc.
%1, 142, 190, and 235
in the test dataset of Defects4J for this example are shown in the
leftmost columns. We expect that {\em the CNN model with its sliding
window is more effective in the resulting matrix after the ordering
due to the nearby dark cells on the left side}. The empirical study on
the impact of such ordering will be explained in Section~\ref{eval}.

\begin{figure}[t]
	\centering
	\lstset{
		numbers=left,
		numberstyle= \tiny,
		keywordstyle= \color{blue!70},
		commentstyle= \color{red!50!green!50!blue!50},
		frame=shadowbox,
		rulesepcolor= \color{red!20!green!20!blue!20},
                xleftmargin=1.5em,xrightmargin=0em, aboveskip=1em,
		framexleftmargin=1.5em,
                numbersep= 5pt,
%		xleftmargin=2.5em,xrightmargin=0em, aboveskip=1em,
%		framexleftmargin=1em,
		language=Java,
 basicstyle=\scriptsize\ttfamily,
    numberstyle=\scriptsize\ttfamily,
    emphstyle=\bfseries,
		%basicstyle=\tiny\ttfamily,
		escapeinside= {(*@}{@*)}
	}
	\begin{lstlisting}[]
public int Compute(int x, int y, int z){
    int i = x + 1;
    int j = x + y;
    int m = 5;
(*@{\color{red}{-\quad if (i < y + 4){}@*)
(*@{\color{cyan}{+\quad if (i < y + 7){}@*)
        if (j > 5 & z > j){
            m = m + z;
        } else {
            m = m + j;
        }
    } else {
        m = m + i;
    }
    i = m + 1;      
    return m;
}
	\end{lstlisting}
        \vspace{-0.12in}
\caption{A Buggy Statement and Interdependent Statements}
\label{fig:motiv_example_II}
%\vskip 4pt
\vspace{-6pt}
\end{figure}

Let us consider another example in
Fig.~\ref{fig:motiv_example_II}. The bug occurs at line 5 and is
fixed in line 6. The program fails in two test cases: 1) \code{x}=5,
\code{y}=0, \code{z}=1, and 2) \code{x}=7, \code{y}=1, \code{z}=9. In
this example, the lines 2, 3, 4, 5, 15, and 16 are all executed in
both passing and failing test cases. Thus, the spectrum-based,
mutation-based approaches, and DeepFL give them the same
suspiciousness scores, and do not detect the buggy line 5 and this
buggy method.
%with the same reasons as explained in the example 1.
%Those lines are equally suspicious with the same suspiciousness
%scores.
The line 16 returns the unexpected results for the two failing test
cases. In fact, the spectrum-based and mutation-based approaches
locate line 16 as the buggy line. However, the actual error occurs at
line 5, steering the execution to the incorrect branch of the \code{if}
statement. This implies that {\em while the source of the bug is at
  line 5, the error exhibits at line 16, which is far apart from
  line 5, yet has a dependency with it}. However, the line 15,
immediate preceding of line 16, does not contribute to the incorrect
result at line 16.

%TIEN: from error mssage link to sus line of code
% not buggy line. Therfore, we need dependencies

\vspace{0.03in}
\noindent {\underline{{\bf Observation 2.}} We observe that the line that
exhibits erroneous behavior (e.g., line 16) might not be the buggy
line (line 5). However, the buggy line 5 has a dependency with the
line 16. Thus, {\em identifying the key line exhibiting the
  erroneous behavior is crucial for FL}.
%\vspace{0.03in}
%\noindent {\underline{{\bf Observation 3.}}
We also observe that the lines with program dependencies with one
another are in fact more valuable in helping localize the buggy line
than the lines without such dependencies. Thus, {\em while considering
the execution order of statements, an FL approach should consider the
statements with program dependencies as well}.

\section{Exploratory Study}
\label{explore}

\begin{figure}[t]
%	\begin{subfigure}{.7\textwidth}
%		\lstset{
%			numbers=left,
%			numberstyle= \tiny,
%			keywordstyle= \color{blue!70},
%			commentstyle= \color{red!50!green!50!blue!50},
%			frame=shadowbox,
%			rulesepcolor= \color{red!20!green!20!blue!20} ,
%			xleftmargin=0em,xrightmargin=0em, aboveskip=1em,
%			framexleftmargin=0em,
%			language=Java,
%			basicstyle=\tiny\ttfamily,
%			escapeinside= {(*@}{@*)}
%		}
%		\begin{lstlisting}[]
%public final void translate(CharSequence input, Writer out) throws IOException {
%	...
%	int pos = 0;
%	int len = input.length();
%	while (pos < len) {
%		int consumed = translate(input, pos, out);
%		if (consumed == 0) {
%			char[] c = Character.toChars(Character.codePointAt(input, pos));
%			out.write(c);
%			pos+= c.length;
%			continue;
%		}
%		for (int pt = 0; pt < consumed; pt++) {
%			(*@{\color{red}{pos += Character.charCount(Character.codePointAt(input, pos));}}@*)
%		}
%	}
%}
%		\end{lstlisting}
%	\end{subfigure}
%	\begin{subfigure}{\textwidth}
		\centering
		\includegraphics[width = 3.6in]{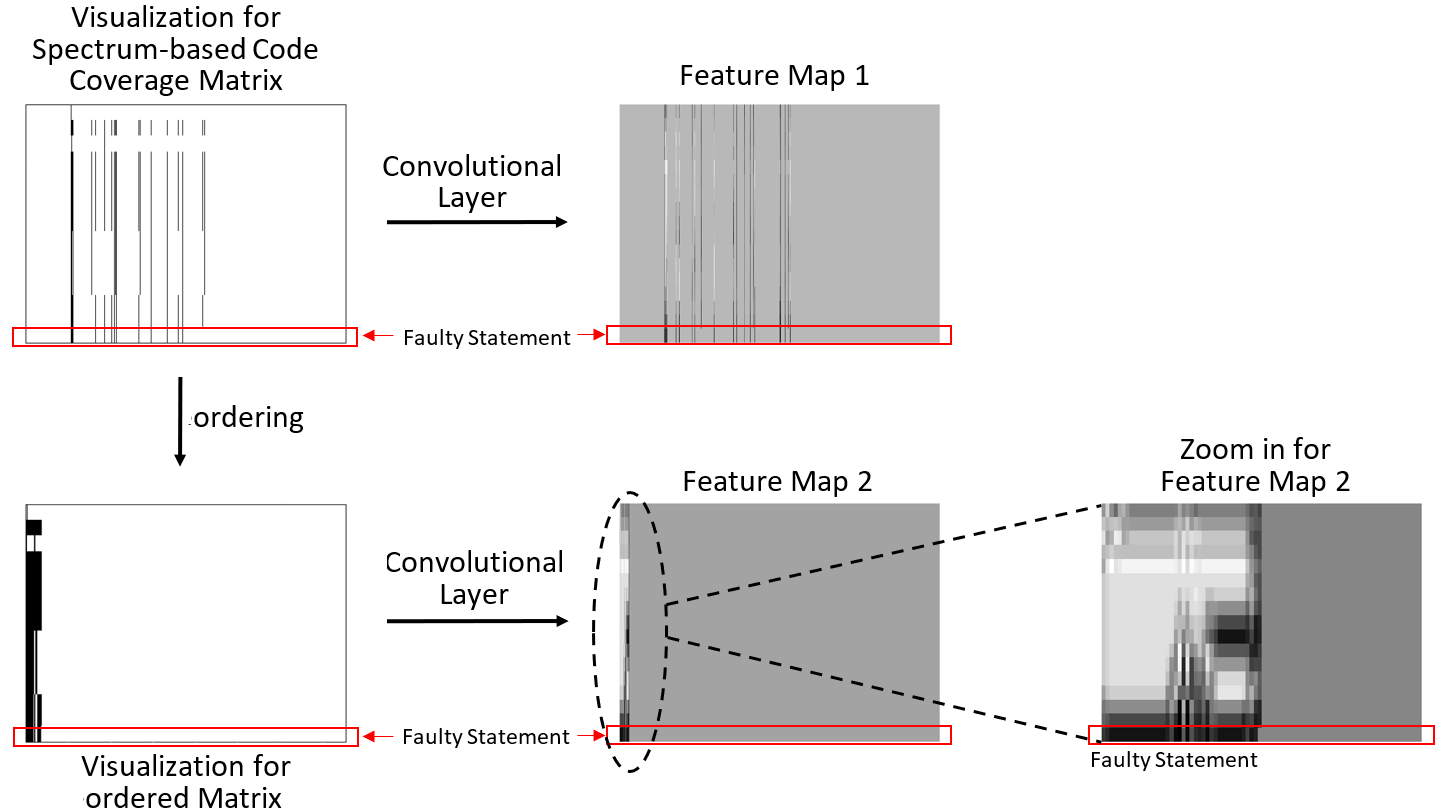}
%	\end{subfigure}
	\caption{A Feature Map after Ordering of Test Cases}
	\label{fig:exploratory}
\end{figure}

%if (out == null) {
%		throw new IllegalArgumentException("The Writer must not be null");
%	}
%	if (input == null) {
%		return;
%	}

Inspiring by the above observations, to further study the impact of
ordering of the columns (i.e, the test cases) of the code coverage
matrix, we conduct an exploratory experiment with the Convolution
Neural Network (CNN) model. Specifically, we choose a simple CNN model
having 2D convolutional layer and 15 convolutional cores with the size
of $3*3$. In this experiment, we use all the bugs in the Defects4J
dataset (will be explained in Table~\ref{fig:data_overview}). As for
training and testing, we use the leave-one-out strategy on the entire
Defects4J dataset. That is, when we perform testing on a fault, we use
all other faults in the dataset as the training data to train the
model. As for the ordering in the code coverage matrix, the first test
case is the one covering as many statements as possible, and the
subsequent test case is the one that runs through as many same
statements as the previously selected test cases (will be detailed in
Section~\ref{1.3}). To encode the pass/fail information, we detected
the error-exhibiting lines (EE) (see Section~\ref{1.2}) and marked
them with -1 values.

%Specifically, the deep learning layer we used to catch the features is the 2D-convolutional layer with 15 convolutional cores with the size of $3*3$ that has been tested as the most suitable number for this experiment. As for training and testing, we use the leaving one out strategy on the whole Defects4J dataset that means when we are doing test on one fault, we use all other faults in the dataset as the training data to train the model. As for the reordered code coverage matrix, we firstly collect error messages from test cases and directly change the value of the crashed statement for the test case to $-1$ in the matrix and then base on some pre-defined rules to reorder the test cases. For the details about adding $-1$ and reordering the test case, you can refer to \textit{section 5.2} and \textit{section 5.3}.

We conduct two executions with two different inputs for the CNN
model. In the first one, for training, we use the original
spectrum-based CC matrix as the input. The output is a matrix with the
same size as the input CC matrix, however, the row corresponding to
the buggy statements/lines are marked with all the 1 values and all
other rows are marked with the zeros. For the second execution, we
use the CC matrix after ordering. The output is the same as in the
first execution.
For testing, we use the trained CNN model to run on the buggy
methods under test. We examine the output of that execution.  The CNN
model generates 15 feature maps as the output from the 15 different
convolutional cores.
The feature maps of a CNN capture the result of applying the CNN
filters to an input matrix. That is, at each layer, the feature map is
the output of that layer. By visualizing a feature map for a specific
input image, i.e., an CC matrix, we aim to gain some understanding of
what features the CNN model can detect.

%let the 2D-convolutional layer generate 15 output as feature maps from
%15 different convolutional cores based on the input. We manually check
%these feature maps to see if our reordering on the code coverage
%matrix can help the CNN learn the key features in the feature maps.

%We setup two models with two different 2D-convolutional layers separately. For training, one uses original spectrum-based code coverage matrix as input and the other one uses reordered spectrum-based code coverage matrix for the method as input. For the output, both of them use a same size matrix as the code coverage matrix that the rows for buggy statements filled with $1$ and other rows filled with $0$. And as for testing, we let the 2D-convolutional layer generate 15 output as feature maps from 15 different convolutional cores based on the input. We manually check these feature maps to see if our reordering on the code coverage matrix can help the CNN learn the key features in the feature maps.

We randomly select 10 faults as the testing data. In two of them,
the result of the CNN model indicates the correct buggy statement for
the fault. Fig.~\ref{fig:exploratory} shows the result for one of the
faults. We visualized the code coverage matrices and feature maps as
gray-scale images. In the code coverage matrices on the left, rows
represent statements from the top to the bottom, and columns represent
test cases. The buggy statement/line is marked with a red rectangle.
As seen, after ordering, the left side of the CC matrix becomes
darker. The white part, which represents the zero values, corresponds
to the test cases that do not go through the statements in this buggy
method.

For the feature maps corresponding to before and after ordering, the
rows also correspond to the statements and the columns represent the
test cases. We examine all 15 feature maps when running the CNN model
on an input. Among the 15 feature maps for the case of ordering, we
found one feature map (feature map 2: the bottom right image) contains
the darker spot at the buggy statement/line compared to the lighter
spots for the non-buggy statements/lines. We examine all 15 feature
maps for the case of the original CC matrix and visualize the
corresponding feature map (feature map 1: the upper right image). As
seen in the red rectangle, there is no dark line/spot around the buggy
statement. In brief, with the ordering of the columns in the CC
matrix, we make the CNN model recognize visual characteristics
corresponding to the buggy statement and distinguish it from the
non-buggy ones. This motivates us to integrate the ordering of
the columns in the CC matrix for code coverage
representation learning.

%to do the testing and 2 of them could locate the fault in the correct
%location. We pick up one from them as an example and show it in figure
%\ref{fig:exploratoty}. As seen in the figure, the red color point out
%which statement is the faulty statement, and for the feature maps
%figure the x-axis is for the test cases and the y-axis is for the
%statements from top to button. Before doing reordering, the
%\textit{feature map 1} learned from the convolutional layer cannot
%catch a good feature for the faulty statement. Just as you can see,
%the information for one statement are spreading in a large range that
%is hard for deep learning model to catch. While after reordering, the
%key information has been grouped together and the left side of the row
%for the buggy statement is darker than all other statements.

%It means that the convolutional layer could catch the key features
%better.

%With the results of this, we come up with our key ideas with a more
%complex model to deal with the fault localization problem with the
%reordering code coverage matrix strategy.

\section{Approach Overview}
\label{keys}

%Inspired by the common method in crime scene investigation, we have
%built our approach, {\tool}, with the key ideas as follows. First, to
%find the perpetrator, an investigator often analyzes the crime scene
%and records what happened. Second, (s)he also looks into the persons
%of interest who are related to the victim with regard to the things
%happened at the scene. Third, (s)he often makes a connection from the
%crime scene to the usual suspects who have committed a similar crime
%with the similar M.O. in the past.

Inspired by the crime scene investigation method, we explore three
aforementioned sources of information. Correspon\-dingly, {\tool} has
three representation learning processes: code coverage
representation learning (crime scene), statements dependency
representation learning (relations), and source code representation
learning (usual suspects) (Fig.~\ref{overview}).

\begin{figure}[t]
	\centering
	\includegraphics[width=3.1in]{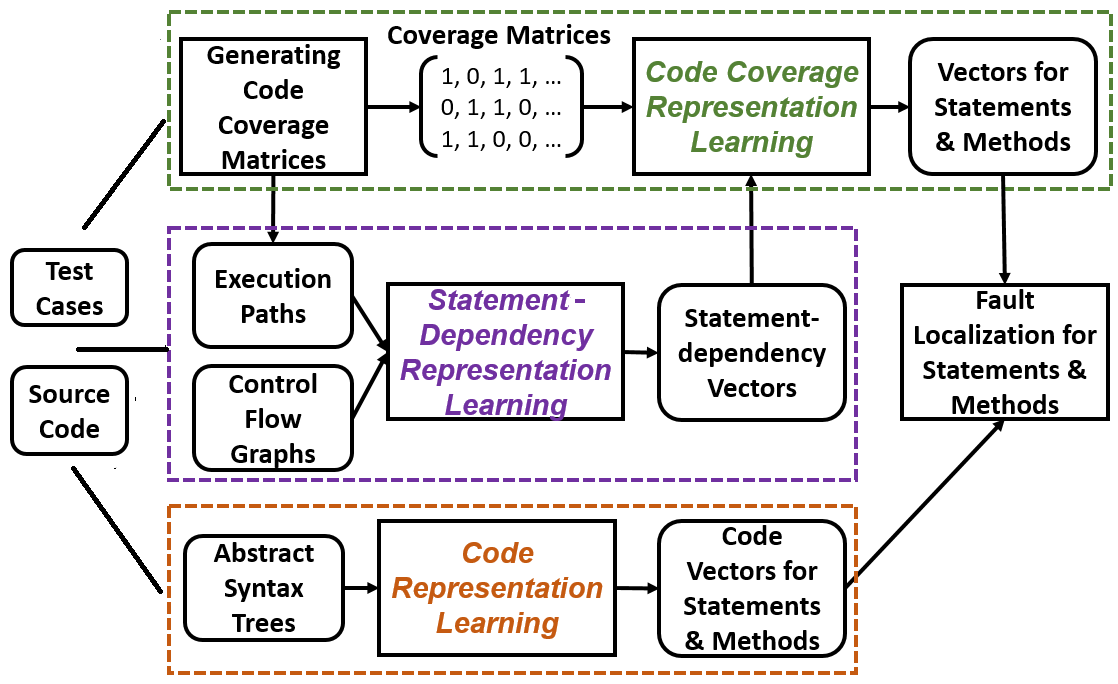}
	%\vspace{-10pt}
	\caption{{\tool}'s Architecture}
	\label{overview}
	\vspace{-8pt}
\end{figure}

\subsubsection{{\bf Code Coverage Representation Learning}}

This learning is dedicated to the ``crime scene'' analysis of the
bug. This process has two parts. First, to help the CNN model
recognize the patterns, we take {\em the given (un-ordered) set of
  test cases} and {\em perform an ordering algorithm} to arrange the
columns of the CC matrix. The strategy of ordering is to enable the
values of 1 to be closer to form darker spots in the left side of the
matrix, expecting that the CNN model can work effectively to recognize
nearby cells to distinguish the buggy and non-buggy statements (see
exploratory study and empirical evaluation).

Second, we also perform the analysis on the output of test cases to
locate the {\em error-exhibiting (EE)} lines (Observation 2). If the
execution of a test crashes, the line information is always
available. Even if there is no crash, the test fails, the program
often explicitly lists the lines of code that exhibit~the incorrect
results/behaviors. We use such information to locate the EE line in
the buggy source code corresponding to each test case. Finally, the
results of individual test cases are encoded as follows.
The cells in the matrix corresponding to the EE~lines of test cases
will be marked with -1 values (see the stars~in
Fig.~\ref{fig:motiv_example_I_c}). Thus, if a column has a value of
-1 at a row, the corresponding test case is a {\em failing} one. The
values of 1 and 0 represent the coverage or non-coverage of the test
case to a statement. Thus, a column has no value of -1 (all the values
are 0 or 1), the corresponding test case is {\em passing}.
%The columns with the values of -1 indicate that~the corresponding
%test cases are the {\em failing} ones, while the columns with the
%values of 1 and 0 represent the {\em passing} ones. The values of 1
%and 0 are for the execution or non-execution.
The resulting matrix is called the {\em enhanced~CC matrix} (ECC).

%In Fig_2, the result of individual test is encoded: if (-1) exists at
%any row, it is a failing test, otherwise passing (i.e.,columns with
%all 0s/1s). (-1) is for error-exhibiting, (1s)/(0s) are for statement
%coverage.

%This matrix is used as the input for our spectrum-based/mutation-based
%representation learning in which the matrix is encoded into the
%vectors to be fed into the CNN model.

\subsubsection{{\bf Dependency Representation Learning}}
%The statements dependency representation learning process is used for
%the investigation of the related statements. That is,
%
The suspiciousness of a statement is seen taking into account the data
dependencies to other statements in the execution flows and data
flows, in addition to the statement itself
(Observation~2). Specifically, we consider both the execution orders
and data dependencies among the statements. For example, if the
statements are executed sequentially in the same nested level as part
of a block statement, data dependencies will help the model in FL as
shown in Section~\ref{motivexample}.
%Tien
%However, if the statements are executed sequentially in the same
%nested level as part of a block statement, we also {\em consider the
%  data dependencies among those statements}.
%  nested within a block statement}.
%Tien removed
%We encode the dependencies among statements in the code representation
%of the ECC as the input to the CNN model to detect the buggy line and
%the buggy method. Together with the representation of all the
%statements in the entire method in the order of appearance in the
%source code, we also encode the data dependencies among the statements
%as explained. In fact,
Additionally encoding the statements with such dependencies has the
same effect as putting together the rows corresponding to the
dependent statements in the CC matrix.
%With this strategy, we expect that the CNN model can characterize
%better the buggy statements via dependencies, and detect the buggy
%line and buggy method as explained earlier.
In our example, in addition to the entire matrix in
Fig.~\ref{fig:motiv_example_II}, we also encode the data
dependencies among statements (i.e., in the same spirit with the case
of putting closer the rows 2, 3, 4, 5, 13, 15, and 16), and feed them
into the CNN model. In our tool, we collect execution paths
and data flow graph for each test case.

\subsubsection{{\bf Source Code Representation Learning}}
For each buggy code in the training data,
%The source code representation learning process is dedicated to the
%part of the investigation considering the usual suspicious code that
%have been buggy in the past.
% From the AST,
we choose to represent the code structure by the long paths that are
adapted from a prior work~\cite{Alon-2018,yioopsla19}. A long path is
a path that starts from a leaf node, ends at another leaf node, and
passes through the root node of the AST. The AST structure can be
captured and represented via the paths with certain lengths across the
AST nodes~\cite{Alon-2018}.
%
%The reason for a path to start and end at leaf nodes is that the leaf
%nodes in an AST are terminal nodes with concrete lexemes.
%The nodes in a path are encoded into a continuous vector. We also
%integrate into our representation the token representation and
%statement representation (Section~\ref{4}). The output of this step
After this, we have the vectors for the buggy code.
Finally, all the representation vectors are used as the inputs
of the CNN model, which is part of the FL module in Fig.~\ref{overview}.

%Tien removed
%Finally, all representation vectors including spectrum based
%representation vectors, mutation-based representation vectors,
%statement dependency representation vectors, and source code vectors
%are combined via Hadamard Product to be used as the inputs of a CNN
%model that acts as the classifier for each statement or method as
%buggy or not. For classification, we use a softmax-based CNN
%classifier. The CNN model is part of the fault localization module
%(Figure~\ref{overview}).

%\section{Approach}

%In this section, we delve into the details of the main steps of our approach.

%\section{\tool:~Deep Representation Learning for Fault Localization}

%\section{Enhancing Code Coverage Matrix}\label{1}

\section{Code Coverage Representation Learning}\label{2}

\subsection{Generating Code Coverage Matrices}\label{1.1}

As in prior FL
studies~\cite{Ochiai,abreu2007accuracy,liblit2005scalable}, we obtain
a code coverage matrix for each method of a given project and error
messages of the failing test cases using GZoltar~\cite{GZoltar}, a
tool for code coverage analysis.  We further modify GZoltar to record
the actual execution path of statements within a method during the
execution of a test case.  For example, for the method in
Fig.~\ref{Fig:motiv_example_I}, the execution path of running the
first selected test case is {\footnotesize $line\_3 \Rightarrow
  line\_6 \Rightarrow line\_7 \Rightarrow line\_(10-11) \Rightarrow
  line\_13 \Rightarrow \underbrace{line\_14 \Rightarrow line\_15
    \Rightarrow line\_17 \Rightarrow line\_18 \Rightarrow
    ... }_{Statements \; repeated \; in \; the \; for \; loop}
  \Rightarrow line\_21$.}

We also use mutation to generate more coverage information. First, we
apply the same mutators as in DeepFL~\cite{DeepFL} to mutate each
statement within a method using the mutation tool
PIT-1.1.5~\cite{PIT}.  To generate a mutation-based matrix, we apply
one mutator to mutate a statement and use GZoltar to record the
execution. Thus, given $n$ mutators that can be applicable to a
statement, we generate $n$ new versions of the given method. If it
has $m$ statements, we generate $n*m$ matrices for the method.  We
refer the mutation-generated $n*m$ matrices as \textbf{mutation-based
  matrices} and for clarification, we refer the non-mutator-generated
matrix as the {\bf spectrum-based matrix}.

%We then apply the same enhancement process as explained earlier to the mutation-based code coverage matrix.

%\vspace{-0.03in}
\subsection{Identifying Error-Exhibiting Lines}\label{1.2}

A cell in the CC matrix can have three values: \{1,0,-1\}. While the
values of 1 and 0 indicate passing, the values of (-1) indicate failing.
We obtain -1 for an error-exhibiting statement or crashed
statement from the error messages of failing test cases. An error
message shows the names of classes, methods, and line numbers
exhibiting an error. We directly use the line numbers, method and
class names to assign -1s to the statements in the matrix.
Fig.~\ref{error} shows an example of the error message containing a
stack trace produced by running a test case on the project {\em Chart}
with the bug {\em Chart-24}. Because the current method under
investigation is \code{getPaint}, our algorithm searches
for that method in the stack trace to derive the EE statement at the
line 128 of the file \code{GrayPaintScale.java} (which contains the method \code{getPaint}).
Each failing test case has only one EE statement for the current method
under study.

\begin{figure}[t]
	\centering
	\includegraphics[width = 3.6in]{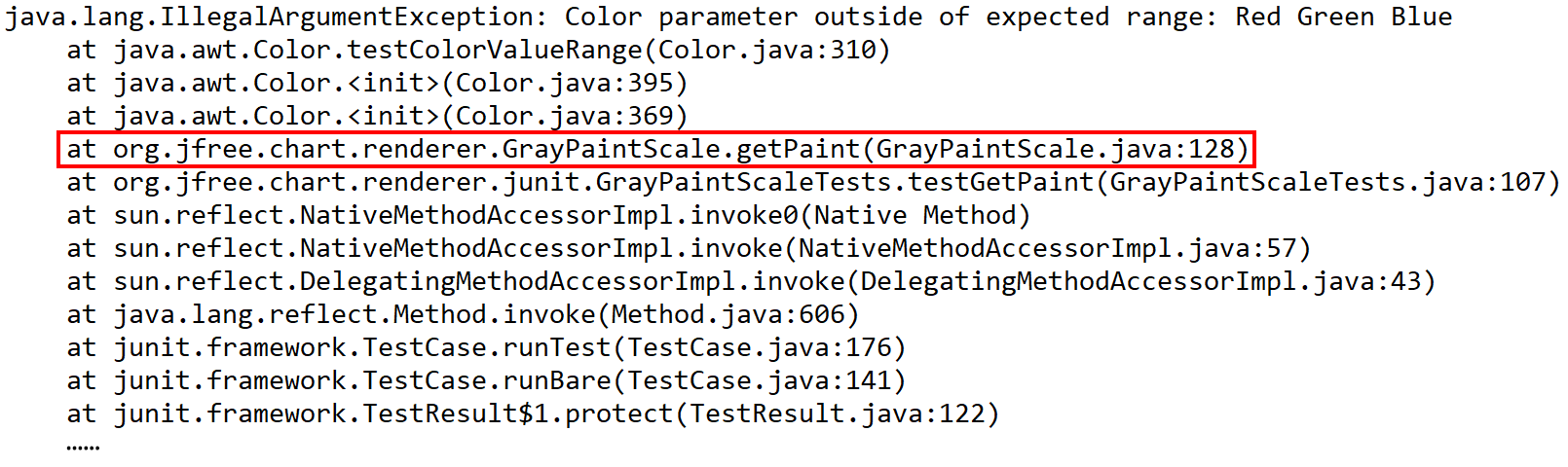}
%		\vspace{-10pt}
	\caption{Error Message Example}
	\label{error} 
%		\vspace{-10pt}
\end{figure}

%\vspace{-0.08in}
\subsection{Test Case Ordering Algorithm}\label{1.3}

Algorithm~\ref{algo} takes the set of test cases $S$ and enforces an
order on $S$. The strategy is to move the values of 1 and -1 closer to
one another in the left side.
%for the CNN model with its small sliding window to work more
%effectively.  To realize our key idea of forming ``visual"
%characteristic features around buggy statements on the left side of
%the matrix, we reorder test cases (i.e., columns) to move \textit{1s}
%and \textit{-1s} closer.  We use the similarity among information of
%cell values: \{~1, 0, -1~\} of columns to order test cases.
%Algorithm~\ref{algo} shows the pseudo-code of our algorithm.
First, if there exist failing test cases, i.e., test cases with -1s,
we select the test case with the value of -1 at the statement
appearing latest in the code.
%
%That is, we use the indexes of the test cases to break the tie (line 5).
We then find the test case that shares the same statement
having -1 with the last selected test case (line 9). That is,
we group together the test cases that go through the
same statement and also fail.
%This will help to form a dark spot in the CC matrix.
If we do not have such test case, then we repeat the process of
looking for another failing test case (i.e., with -1).  In
Fig.~\ref{fig:motiv_example_I_c}, the test case 9 is selected as the
first one with only one -1 (marked with a star) at the line 13 (latest
statement). We search for the next test case that has a -1 at the
latest. The test case 33 is chosen at the second column.

{\small
\begin{algorithm}[t]
\caption{Test Case Ordering Algorithm}
 \label{algo}
 \scriptsize
\begin{algorithmic}[1]
\Function {OrderingTestCases}{$S: testSet$}
\State $List = []$
\While {$(S <> \emptyset)$}
	\If {$\textsc{HaveTestCasesWithMinusOne}(S)$}
            \State $selT = \textsc{FindTestCaseWithMinusOneWithHighestIndex}(S)$
		\State $S.remove(selT)$
		\State $List.append(selT)$
		\While {$\textsc{HaveTestCaseSameStmtWithMinusOne}(selT, S)$}
			\State $selT = \textsc{FindTestCaseSameStmtWithMinusOne}(selT, S)$
			\State $S.remove(selT)$
			\State $List.append(selT)$
		\EndWhile
          \Else
                \State $selT = \textsc{FindTestCaseWithMostOne}(S)$
		\State $S.remove(selT)$
		\State $List.append(selT)$
		\While {$\textsc{HaveTestCaseWithSameStmtsWithOne}(selT, S)$}
			\State $selT = \textsc{FindTestWithMostSameStmtsWithOne}(selT, S)$
			\State $S.remove(selT)$
			\State $List.append(selT)$
		\EndWhile
	\EndIf
\EndWhile
\State \textbf{return} $List$

\if\else
\State $stepOne = True$
\State $reorderedCodeCovMatrix = NewMatrix()$
\State $lastTestCase = empty$
%\if\else
\For {\textbf{each} $(testCaseNum \in Length(codeCovMatrix))$}
	\For {\textbf{each} $(testCase \in codeCovMatrix)$}
		\If {$stepOne$ $\textbf{\&}$ $HasMostOneOrMinusOne(testCase, codeCovMatrix)$ }
			\State $reorderedCodeCovMatrix.append(testCase)$
			\State $stepOne = False$ 
			\State $codeCovMatrix.pop(testCase)$
			\State $lastAddedTestCase = testCase$
			\State $Break$
		\Else {
		\If {$!stepOne$ $\textbf{\&}$ $HasMostSameState(testCase, lastTestCase, codeCovMatrix)$}
			\State $reorderedCodeCovMatrix.append(testCase)$
			\State $stepOne = False$ 
			\State $codeCovMatrix.pop(testCase)$
			\State $lastAddedTestCase = testCase$
			\If {$!stepOne$ $\textbf{\&}$ $!HasSameState(lastTestCase, codeCovMatrix)$}
				\State $stepOne = True$
			\Else {  \textbf{continue}}
			\EndIf
		\EndIf}
		\EndIf
	\EndFor
\EndFor

\State \textbf{return} $reorderedCodeCovMatrix$
\fi
\EndFunction
%\if\else

\if\else
\Function {HasMostOneOrMinusOne} {$testCase$, $codeCovMatrix$}
\If {$HasMinusOne(codeCovMatrix)$}
	\For {\textbf{each} $(testCaseSearch \in codeCovMatrix)$}
		\If {$NumberOfMinusOne(testCaseSearch) > NumberOfMinusOne(testCase)$}
			\State \textbf{return} $False$
		\Else {  \textbf{continue}}
		\EndIf
	\EndFor
\Else {
\If {$!HasMinusOne(codeCovMatrix)$}
	\For {\textbf{each} $(testCaseSearch \in codeCovMatrix)$}
		\If {$NumberOfOne(testCaseSearch) > NumberOfOne(testCase)$}
			\State \textbf{return} $False$
		\Else {  \textbf{continue}}
		\EndIf
	\EndFor
\EndIf }
\EndIf
\State \textbf{return} $True$
\EndFunction

\\

\Function {HasMostSameState} {$testCase$, $lastTestCase$, $codeCovMatrix$}
\If {$HasMinusOne(lastAddedTestCase)$}
	\For {\textbf{each} $(testCaseSearch \in codeCovMatrix)$}
		\If {$SameMinusOne(testCaseSearch, lastTestCase)$ $\textbf{>}$ $SameMinusOne(testCase, lastTestCase)$}
			\State \textbf{return} $False$
		\Else {  \textbf{continue}}
		\EndIf
	\EndFor
\Else {
\If {$!HasMinusOne(lastAddedTestCase)$}
	\For {\textbf{each} $(testCaseSearch \in codeCovMatrix)$}
		\If {$SameOne(testCaseSearch, lastTestCase)$ $\textbf{>}$ $SameOne(testCase, lastTestCase)$}
			\State \textbf{return} $False$
		\Else {  \textbf{continue}}
		\EndIf
	\EndFor
\EndIf }
\EndIf
\State \textbf{return} $True$
\EndFunction

\\

\Function {HasSameState} {$lastTestCase$, $codeCovMatrix$}
\For {\textbf{each} $(testCase \in codeCovMatrix)$}
	\If {$HasSameStateWithOneOrMinusOne(lastTestCase, testCase)$}
		\State \textbf{return} $True$
	\Else {  \textbf{continue}}
	\EndIf
\EndFor
\State \textbf{return} $False$
\EndFunction
\fi
\end{algorithmic}
\end{algorithm}
}

If we do not have any failing test case left, we select the test case
that has the most 1s (line 13).
%For each test case, we count the number of \textit{1s} and
%\textit{-1s} of the column. We select the test case that has the most
%\textit{1s} (line 5). However, if multiple test cases have the same
%number of \textit{-1s} with the last selected column, we choose the
%one with the most number of \textit{1s} (line 13); if multiple test
%cases still have the same number of both \textit{-1s} and \textit{1s},
%we selected the one with the higher index in the test suite. That is,
%we use the indexes of the test cases to break the tie.
Next, we select the next test case that shares the most number of the
same statements having the values of 1s with the last
selected test case. This helps move the values of 1 closer.
%This will also help to form a dark spot in the left side of the CC
%matrix.
%
We repeat this step to select a new test case compared with the
previously selected one until all the test cases were ordered. We stop
this step if no test case has the same statements with 1s as the last
selected test case (column).
%Finally, the columns with all the zeros are placed according to their
%indexes in the test suite from the highest to the lowest.
If two test cases are tie, we select the one with the last value of 1
at a statement appearing latter. The rationale is that such a test
case covers more statements than the other. If they are still tie, the
selection of either of them will result in similar visual effects
locally at that~row. In brief, in any cases of ties, the visual
effects around the statements are similar.

%In addition to all previously selected columns, we re-start the first
%step and repeat the second step.  If all columns having \textit{-1s}
%are ordered, we conduct the same ordering procedure for the other
%columns that have only \textit{1s} with \textit{0s}. Then, we randomly
%order the remaining columns with \textit{0s}.
%------------------------------------------------------------------

%Tien removed
%{\bf OLD.} After these steps, we obtain the enhanced code coverage
%matrix (ECC). For example, the ECC matrix in
%Figure~\ref{fig:motiv_example_II_c} can show the information for both
%code coverage and test case results for each line using \textit{0, 1,
%  and -1}. For example, \textit{line 16} has the \textit{-1} for the
%test cases 21 and 33 to indicate that (1) these two test cases execute
%the \textit{line 16} but have errors at this statement; and (2) test
%cases 21 and 33 have relations and they are dependent.

In addition to the spectrum-based matrices, we also apply the same
enhancements, \textit{identifying error-exhibiting lines} and
\textit{ordering text cases} to  mutation-based code coverage
matrices.

%----- MOVED

%\input{sections/combine}

\section{Statement-Dependency Representation}\label{3}

%In this step,

%\subsection{Dependencies Representation Learning}
%Given

We aim to model the {\em execution orders} and {\em data dependencies}
among the statements of the method under study.

%using the statement execution paths and data-flow graphs.

\subsubsection{{\bf Execution Order Representation}}
We obtain the execution path (e-path) as each test case was
executed. We only consider the relations among statements within a
method.
%, thus we remove any method call links when generating an
%execution path of statements within a method for a test case.
%
%But with the same reason as $G_1$, we only want to consider the relationship inside of a method. So we modify the execution paths we have, we split them by method and then we can have have a small execution path $EP_u$ for each method $F_u$. When doing it, if there is a method call happens in the method, we will cut the running process in other methods and link the previous and next statements in the method directly in order to keep the execution path works right.
%We model the relations between a statement and its neighboring statements in an E-path.
Since an e-path is a sequence of statements, we apply
{\em word2vec}~\cite{word2vec} on all execution paths of test cases to learn
the vectors that encode the relations among statements.
%, where we treat each statement as a ``word" and an e-path as a
%``sentence".  Word2Vec is a neural network that takes a text corpus as
%an input and generates a set of feature vectors for words in the
%corpus.
Thus, \textit{each statement has a word2vec-generated vector}.

\subsubsection{{\bf Data Dependency Representation}}
\label{dpsection}

Using execution paths is not sufficient due to the following.
%Although execution paths can capture some dependency relations among
%statements, it is not sufficient due to the following reasons.
First, the statements in a loop may repeat multiple times in an e-path,
thus, they 
%the relations among these statements in a loop
may dominate vector learning using {\em word2vec} and weaken the
relations between the statements inside and outside of a loop, which
is also crucial in FL.
%
%However, quite often the dependency between the statements
%inside and outside a loop can help localize bugs.
%
Second, interdependent statements might not be nearby in an e-path,
yet are useful in detecting the buggy line (Observation 2). To
address those, we also use the data-flow graph (DFG) for the
statements in a method.
%, in addition to the execution paths of test cases.

% Tien removed
%\begin{figure}[t]
%	\centering
%	\includegraphics[width = 3.4in]{graphs/Node2vec.png}
%	\caption{The partial DFG of Variable $m$ in Figure~\ref{fig:motiv_example_II}}
%	\label{n2v} 
%\end{figure}

We use WALA~\cite{WALA} to generate DFGs in which a node~represents a
statement and an edge represents a data flow between two nodes. If $A$
connects to $B$, we assign the weight of~1. If there is no edge from
$B$ to $A$, we create that edge but assign the weight of -1. This
makes {\em node2vec}~\cite{Grover-2016}, a network embedding technique,
applicable to our graph. The value of -1 helps distinguish between the
artificial edges and the real ones.  After this step, some statements
(nodes) with data dependencies have {\em node2vec}-generated vectors.

%Tien removed
%We aim to capture the dependency relations among nodes into a vector
%representation that is required by the CNN model. Specifically, we
%first modify a (unidirectional) DFG to be bi-directional, as {\em
%  node2vec}~\cite{Grover-2016}, a widely used network embedding
%technique, is applicable to a network that supports two-way
%information flow. Between two nodes (e.g., $N_i$ and $N_j$), we assign
%1 to the edge from $N_i$ to $N_j$ to indicate the actual flow of data,
%and create an opposite edge from $N_j$ to $N_i$ with a weight -1 to
%indicate there should be no flow of data.  After the modification,
%each pair of nodes has two edges. We directly apply {\em
%  node2vec}~\cite{Grover-2016} on the modified DFGs to generate node
%embeddings.
%Figure~\ref{n2v} shows the key steps of this process.

%Tien removed
%In Figure~\ref{n2v}, for simplicity, we use a partial DFG
%of the method in Figure~\ref{fig:motiv_example_II}, regarding only the
%variable $m$ as an example (a complete DFG also contains data flows
%for all variables).  We modify the partial DFG
%regarding the variable $m$ into bi-directional. Then we use {\em
%  node2vec} as a black box tool to take the modified DFG, and
%output the vectors for all nodes in the DFG.  After this step, {\it some
%  statements with data dependencies have node2vec-generated vectors.}

\subsubsection{{\bf Vectors for Statements with Dependencies}}\label{3.3}
%If a statement is not in any block, we use its {\em
%  word2vec}-generated vector as the final one, otherwise the {\em
%  node2vec}-generated vector is chosen for the statement.
%Word2vec for ***statement order*** and node2vec for ***dependencies***  are combined via Hadamard product to be the statement vector. Top-1 is improved from 226 to 245 with statement relations (both order and dependencies)(Table_IV). We’ll measure the contribution only by statement order.
The {\em word2vec} vector for a statement $s$ in the execution order and the
{\em node2vec} vector for $s$ in program dependencies among the statements
are combined via Hadamard product to represent $s$.
Finally, the output vector is {\bf a statement-dependency~vector for
  a statement}, modeling the statement with the dependencies and/or
execution orders among statements.

\subsubsection{{\bf Combining Statement Dependencies and ECC Matrices}}\label{2.1}

To further enrich the ECC matrix (a spectrum-/mutation-based matrix),
we incorporate the dependencies among the statements in a method under
study into that matrix.
%
%We build statement-dependency vectors to model the dependencies among
%the statements in a method. Each statement has a 1-dimensional
%statement-dependency vector. (We will explain how we model the
%dependencies among the statements in Section~\ref{3}).
%
%Let us explain how we add the statement-dependency vectors into the
%ECC matrix.
In the enhanced matrix, we have the $i$-th statement ($S_i$) of a
method under test with the test cases, $T=\{T_1$, $\dots$, $T_j$
$\dots$, $T_n\}$, where $j$ indicates the $j$-th test case, $1 \le j
\le n$, and $n$ is the number of test cases. The statement $S_i$ under
a test case $T_j$ has a cell value $v_{ij}$ that can be either \{1, 0,
or -1\}.  Thus, the statement $S_i$ can be represented as a vector
$\vec{S}_i = \{v_{i1}, \dots, v_{ij}, \dots, v_{in}\}$.  Each
statement ($S_i$) has a statement-dependency vector
($\vec{S}^{sd}_i$).  We multiply each $v_{ij}$ with $\vec{S}^{sd}_i$,
to obtain $v_{ij}*\vec{S}^{sd}_i$, for each cell of $S_i$ and $T_j$ in the
enhanced matrix.  Thus, the statement $S_i$ can be represented as a
new 2-dimensional vector $\vec{S}^{2d}_i$ = $<v_{i1}*\vec{S}^{sd}_i, \dots,
v_{ij}*\vec{S}^{sd}_i, \dots, v_{in}*\vec{S}^{sd}_i>$.  Any vector $\vec{S}^{sd}_i$
multiplied by a $v_{ij}=0$ results in a vector with all 0s.

%For example,
%a statement \textit{line-16} can be represented as a list of
%statements-dependency vectors: <1*<$V_{S_{16}}$>, 0*<$V_{S_{16}}$>,
%... , 1*<$V_{S_{16}}$> , ... , 1*<$V_{S_{16}}$>, ...>, a 2-dimensional
%(D) matrix. The method can be represented as a list of statements and
%each statement as a 2-D matrix, so the method as a 3-D matrix.

A method often has multiple statements $\{S_1$,$\dots$, $S_i$, $\dots$
$S_m\}$, where $i$ indicates the $i$-th statement, $1 \le i \le m$, and
$m$ is the number of statements.  Thus, {\em a method is presented as
a 3-D matrix, i.e., a list of 2-D statement vectors}.

The same steps are used to enhance and combine statement dependencies
into a mutation-based matrix.  A statement $S_i$ in a mutation-based
matrix is represented as a set of mutated statements and each mutated
statement is represented as a 2-D vector.  Thus, in this case, the
statement $S_i$ is represented as a 3-D matrix. After enhancing the
ECC matrix and combining statement-dependencies as explained, we
obtain the following:
\begin{itemize}
	\item In a spectrum-based matrix (SBM), a statement is represented
	as a 2-D vector and a method as a 3-D matrix;
	\item In mutation-based matrices (MBM), a statement is represented
	as a 3-D matrix and a method as a 4-D matrix.
\end{itemize}

\subsubsection{{\bf Encoding Code Coverage Matrices with a CNN Model}}\label{2.2}
%After enriching the CC matrices, we obtain the above mentioned
%representations for statements and methods.
%Each of these representations is a matrix, thus it is natural to apply
After obtaining those representations for statements and methods, we
apply the Convolution Neural Network (CNN)~\cite{kim2014convolutional}
to learn features. We use a typical CNN with the following layers: a
convolutional layer, a pooling layer and a fully connected layer.  We
feed the followings into the CNN model separately to detect a buggy
statement/method:

\noindent i) For spectrum-based matrices (SBM), we fed a 2-D vector representing for a statement and a 3-D matrix for a method,

\noindent ii) For mutation-based matrices (MBM), we fed a 3-D matrix representing for a statement and a 4-D matrix for a method.

%(1) a statement represented as a 2-D vector in a spectrum-based matrix (SBM);

%(2) a method represented as a 3-D vector in a SBM;

%(3) a statement represented as a 3-D vector in a mutation-based matrix (MBM); and

%(4) a method represented as a 4-D vector in a MBM,

We apply a fully connected layer before CNN on the method in a
mutation-based matrix (i.e., represented as a 4-D matrix) to reduce
an 4-D matrix into an 3-D matrix.

The outputs of the CNN include the vectors for a statement or a
method in spectrum-based or mutation-based matrices:

i) $V_{ss}$, 1-D vector for a statement in SBM,

ii) $V_{sm}$, 1-D vector for a method in SBM,

iii) $V_{ms}$, 1-D vector for a statement in MBM, and

iv) $V_{mm}$, 1-D vector for a method in~MBM.

\section{Source Code Representation Learning}\label{4}

%After learning the test case coverage information, in order to do more detailed analysis on code, we would like to do the code representation learning in this step. To do it, we mainly do the analysis on the relationship among tokens and the relationship among statements. We expect the source code as the input of this step. And then the output would be statement/method representation vectors. Because the way of learning representation vectors are different in method level and statement level in this step, we talk about them on both statement level and method level here. To introduce them, we would like to talk about them in two small steps.

%We target at fault localization at two levels: statement and method. However, the representations used for each level are different, thus, we introduce them separately.

Let us explain how we capture the usual suspicious source code via
code representation learning.

For a statement, we tokenize it and treat each token in the
statement as a word and the entire statement as a sentence.  We use
{\em word2vec}~\cite{word2vec} on all the statements of a project to
compute a token vector for each token.
%For example, in Figure~\ref{fig:motiv_example_II}, the buggy line
%\textit{if (i < y + 4 )} is treated as a sentence with its tokens
%\textit{"if", "(", "i", "<", "y", "+", "4", ")"} as words.
After having the vectors for all the tokens, for a
statement, we have a matrix [Token-Vector$_1$, Token-Vector$_2$,
  $\dots$, Token-Vector$_m$]. To obtain a unified vector to
represent a statement instead of a matrix, we apply a fully connected
layer to reduce the matrix into 1-D vector. Thus, we have one vector
for each statement.

%Tien Removed
%\begin{figure}[t]
%	\centering
%	\includegraphics[scale=0.17]{graphs/AST.png}
%	\vspace{-10pt}
%	\caption{Example of Long-Path (Red) and Sub-Tree (Green)}
%	\label{fig:subtree_longpath}
%	\vspace{-10pt}
%\end{figure}

\iffalse
Collectively, We run word2Vec on a large corpora for all statements to learn a vector $TokenV_i$ to represent a token $n_i$ in a statement $S={n_1, n_2,..., n_i}$. All tokens in training are considered to maximize the log value of the probability of neighboring tokens in the input data set. We use word2vec to train our own tokens representations by using all of the tokens from each project.
	
The loss function is defined as follows:
\vspace{2pt}
\begin{equation} \label{eq:1}
\resizebox{.38 \textwidth}{!}
{
	$L(i) =\min_i{\frac{1}{i}\sum_{j=1}^i \sum_{k \in NNS} -\log HS\{TokenV_k|TokenV_j\}}$
}
\vspace{2pt}
\end{equation}
where $L$ is the lose function for the tokens in $S={n_1, n_2,...,n_i}$, $NNS$ is the set of the neighboring tokens of a token $n_i$, and $HS\{TokenV_k|TokenV_j\}$ is the hierarchical softmax of token vectors $TokenV_k$ for the node $n_k$ and $TokenV_j$ for the node $n_j$.
\fi
	
%\subsection{Method-level Code Representation Learning} \label{4.2}

At the method level, we used two existing code representation learning
techniques {\em code2vec}~\cite{yioopsla19} and
ASTNN~\cite{zhang2019novel} for a method. In {\em code2vec}, we
use {\em long paths} over the AST. A long path is
a path that starts from a leaf node, ends at another leaf node, and
passes through the root node of the AST.
%(e.g., the red dash path in Figure~\ref{fig:subtree_longpath}).
The AST structure can be represented via the paths with certain
lengths across the AST nodes.
%
%We used the same technique in {\em code2vec}~\cite{yioopsla19} to
%build long paths.
Specifically, we regard a long path as a sequence and apply {\em
  word2vec} on all the long paths of a method to generate a vector
representation for each AST node. Now, each path is represented as
an ordered list of node vectors (the order is based on the appearance
order of the nodes in a path), and each method is represented as
a bag of paths, i.e., ordered lists of node vectors. Essentially, a
method is represented by a matrix. We use a fully connected layer to
transform the matrix into 1-D vector for a~method.

At the method level, we also used tree-based representation
ASTNN~\cite{zhang2019novel}. ASTNN splits
the AST of a method into small subtrees at the statement level and applies a
Recursive Neural Network (RNN)~\cite{socher2011parsing} to learn
vector representations for statements. The ASTNN exploits the
bidirectional Gated Recurrent Unit~(GRU)~\cite{tang2015document} to
model the statements using the sequences of sub-tree vectors.
%For example, the green box in Figure~\ref{fig:subtree_longpath}
%indicates a sub-tree for the whole AST.
%
After obtaining the long-path-based vector and the tree-based vector
for a method, we apply a fully connected layer as the one in
CNN~\cite{kim2014convolutional} to combine these two 
vectors into one unified vector for a method.

\section{Fault Localization with CNN Model}

%In the previous sections, {\tool} learns vectors for statements
%and methods from the enhanced matrices and source code to perform
%fault localization at statement- and method levels.

\subsection{Statement-level Fault Localization}\label{stm-level}
After all the previous steps, each statement has 3 vectors:

\noindent 1) $\vec{V}_{ss}$, a SBM-based statement vector (Section~\ref{2.2});

\noindent 2) $\vec{V}_{ms}$, a MBM-based statement vector (Section~\ref{2.2}); and

\noindent 3) $\vec{V}_{cs}$, a source code-based statement vector (Section~\ref{4}). 

%We combine the above three vectors to obtain a matrix for a statement
%by vector multiplication.

\noindent The vectors are combined via Hadamard Product~\cite{hadamard}:
{\small
$$M_s = [len(\vec{V}_{ss}), 1, 1], M_m = [1, len(\vec{V}_{ms}), 1], M_c = [1, 1, len(\vec{V}_{cs})]$$
$$M_{combined} = broadcast(M_s) \circ broadcast(M_m) \circ broadcast(M_c)$$
}
$M$ is the matrix which is expanded from $v$ by keeping one dimension
as $v$ and adding two more dimensions with the size of $1$.
%$v$ is in the threes vectors we mentioned above; $\circ$ is Hadamard
%product;
$broadcast()$ is the operation to copy a dimension into multiple times
to expand the matrix to the suitable size for Hadamard product. The
rationale is that all three vectors from three different aspects
should be fully integrated. The resulting matrix is of the size
$[len(\vec{V}_{ss}), len(\vec{V}_{ms}), len(\vec{V}_{cs})]$.
Next, we use the trained CNN model with a softmax on the matrix to
classify a statement into faulty or non-faulty. The output of the
softmax is standardized to be between $0$ to $1$. To train
the model, the same combined matrix for a statement is used at the
input layer and the corresponding classification
(faulty or not) is used at the output layer of the CNN~model.

%With the result of this, we can setup a threshold to do the fault localization. 

\subsection{Method-level Fault Localization}

Similar to statement-level FL, each method has 3 vectors:

\noindent 1) $\vec{V}_{sm}$, a SBM-based method vector (Section~\ref{2.2});

\noindent 2) $\vec{V}_{mm}$, a MBM-based method vector (Section~\ref{2.2}); and

\noindent 3) $\vec{V}_{cm}$, a source code-based method vector
(Section~\ref{4}).

Moreover, we also consider the similarity between the source code and
the error messages of the failing test cases as in
DeepFL~\cite{DeepFL}.
%For fault localization at the method level, we also consider
%the relations between the source code of the method and the error
%messages of failing test cases for the method-level fault
%localization. DeepFL~\cite{DeepFL} shows that
%calculating the similarities between methods and error messages of
%failing test cases can improve method-level fault localization. We
%adopt that technique.
We first collect 3 types of information from failed tests, including
the name of the failed tests, the source code of the failed tests and
the complete failure messages (including exception type, message, and
stacktrace). Second, we collect 5 types of information from source
code, including the full qualified name of the method, accessed
classes, method invocations, used variables, and comments.  For each
combination, we calculate the similarity score between each
information from the failed tests and each from the source code using
the popular TF-IDF method~\cite{DeepFL}. We generate 15 similarity
scores as 15 features for a method. Thus, a method also has the fourth
vector, $\vec{V}^{sim}_m$ with 15 features.

%Thus, a method can be represented as a vector with 15 feature values,
%$\vec{V}^{sim}_m$.  After this, each method has 4 vectors:

%(1) $\vec{V}^{sc}_m$, an enriched spectrum-based coverage method vector;

%(2) $\vec{V}^{mc}_m$, an enriched mutation-based coverage method vector;

%(3) $\vec{V}^{s}_m$, a source code method vector; and (4) $\vec{V}^{sim}_m$.

For fault localization, we combine the above method vectors into a matrix by using the Hadamard product as in Section~\ref{stm-level}, 
then use the trained CNN model
with a softmax to classify a method into faulty or non-faulty. We
train the model in the same manner as FL at the statement level.

\section{Empirical Evaluation}
\label{eval}
%\vspace{-0.06in}

%In this section, we introduce our research questions, analysis approaches, and empirical results.

%We conducted several experiments to evaluate {\tool}.

\subsection{Research Questions}

%We have conducted several experiments to evaluate our model. Specifically, 
%Through empirical studies, 
We seek to answer the following research questions:

\noindent\textbf{RQ1. Statement-level FL Comparison.} How well does our tool  perform compared with
the state-of-the-art {\em statement-level} FL models?

%at the statement level with the existing state-of-the-art fault localization approaches?

\noindent\textbf{RQ2. Method-level FL Comparison.} How well
does our~tool perform compared with the existing {\em method-level} FL models?

%\noindent\textbf{RQ3. Advanced Model Application Comparative Study.} Can using the advanced deep learning model

%that helpful for improving the performance of the whole model by using the advanced deep learning model?

%\noindent\textbf{RQ3. Sensitivity Analysis.} How do various factors affect the overall performance of {\tool}?

\noindent\textbf{RQ3. Impact Analysis of Different Matrix Enhancing
  Techniques.}  How do those techniques including test case ordering,
and statements dependency affect the accuracy?

\noindent\textbf{RQ4. Impact Analysis of Different Representations
  Learning.} How do different types of information affect
the accuracy?

%How do different types of representation for the methods/statements affect the overall performance of {\tool}?

%\noindent\textbf{RQ5. Impact Analysis of Statements Dependency and Textural Similarity between Source Code and Error Messages.} How do statements dependency and textural similarity between source code and error messages affect the overall performance of {\tool}?

%\noindent\textbf{RQ4. Advanced Models Analysis} Can advanced deep learning models improve our {\tool}?

%that helpful for improving the performance of the whole model by using the advanced deep learning model?

\noindent\textbf{RQ5. Cross-project Analysis.} How does {\tool} perform in the cross-project setting?

\noindent\textbf{RQ6. Performance on C Code.} How does {\tool} perform in C projects for FL?

%\noindent\textbf{RQ5. Impact Analysis of Statements Dependency and Textural Similarity between Source Code and Error Messages.} How do statements dependency and textural similarity between source code and error messages affect the overall performance of {\tool}?

%\noindent\textbf{RQ7. Sensitivity Analysis.} How do various factors affect the overall performance of {\tool}?

\subsection{Experimental Methodology}

\subsubsection{\textbf{Data Set}}
We use the benchmark, Defects4J V1.2.0~\cite{defects4j} with
ground truth (Table~\ref{fig:data_overview}).
For a bug in project $P$, Defects4J has a separate copy of $P$ but
with only the corresponding test suite revealing the bug. For example,
$P_1$, a version of $P$, passes a test suite $T_1$. Later, a bug $B_1$
in $P_1$ is identified. After debugging, $P_1$ has an evolved test
suite $T_2$ detecting the bug. In this case, Defects4J has a separate
copy of the buggy $P_1$ with a single bug, together with the test
suite $T_2$. Similarly, for bug $B_2$, Defects4J has a copy of $P_2$
together with $T_3$ (evolving from $T_2$), and so on.
%We do not use the whole T of all test suites for training/testing.
For within-project setting, we test one bug $B_i$ with test suite
$T_{(i+1)}$ by training on all other bugs in $P$.
%
%We use all of the 6 projects in Defects4J V1.2.0 with 395 real bugs
%(Table~\ref{fig:data_overview}). In the dataset, each bug contains a
%buggy version of project that includes a large number of training
%instances (i.e., method/statements). For example, the project Math has
%140,000+ training instances.
To reduce the influence of the overfitting problem, we applied L2
regularization and added dropout layers.

%shows the statistics of Defects4J V1.2.0.

\begin{table}
%	\vspace{-8pt}
	\caption{Defects4J Dataset}
	\vspace{-6pt}
	{\small
	\begin{center}
		\renewcommand{\arraystretch}{1}
		\begin{tabular}{l|ll}
			\hline
			Identifier & Project name & \# of bugs \\
			\hline
			Chart & JFreeChart & 26 \\
			Closure & Closure compiler & 133 \\
			Lang & Apache commons-lang & 65 \\
			Math & Apache commons-math & 106 \\
			Mockito & Mockito & 38 \\
			Time& Joda-Time & 27 \\
			\hline
		\end{tabular}
		\label{fig:data_overview}
	\end{center}
    }
\vspace{-14pt}
\end{table}

%, including 200+ unique dynamic or static fault localization features
%for each program element within each buggy version.

\subsubsection{\textbf{Experiment Metrics}}

Following prior studies~\cite{DeepFL,TraPT}, we use the following
metrics to evaluate an FL model:

\textbf{Recall at Top-K:} is the number of faults with at least one
faulty statement that is correctly predicted in the ranked list of $K$
statements. We report Top-1, Top-3, and Top-5.

%In practice, a developer focuses on inspecting the top-ranked buggy
%statements/methods during fault localization. Too many results may
%confuse and waste developers' efforts in fault localization.
%Therefore, we use Top-1 as the key metric to evaluate {\tool} and the
%baselines. In addition, we also provide the results for Recall at
%Top-3 and Top-5.

\textbf{Mean Average Rank (MAR):} We compute the average rank of
all faulty elements for each fault. MAR of each project is
the mean of the average rank of all of its faults.

\textbf{Mean First Rank (MFR):} For a fault with multiple faulty
elements (methods/statements), locating the first one is
critical since the others may be located after that. MFR of each
project is the mean of the first faulty element's rank for each fault.

\vspace{0.03in}
\subsubsection{\textbf{Experiment Setup and Procedure}}~\\\noindent\textbf{RQ1: Statement-level Fault Localization Comparison.}

\textit{Baselines.}
We compare {\tool} with the following statement-level FL approaches:
\begin{itemize}
	\item Two spectrum-based
	fault localization (SBFL) techniques: \textbf{Ochiai}~\cite{Ochiai}
	and \textbf{Dstar}~\cite{DStar};
	\item Two recent mutation-based fault
	localization (MBFL) techniques: \textbf{MUSE}~\cite{MUSE} and
	\textbf{Metallaxis}~\cite{Metallaxis};

	\item
	Two deep-learning based FL approaches: \textbf{RBF Neural
          Network (RBF)}~\cite{RBF_Neural_Network} and
        \textbf{DeepFL}~\cite{DeepFL}.  DeepFL~\cite{DeepFL} works at
        the {\em method level} with several features. For comparison,
        in this RQ1 for the statement level, we can only use
        DeepFL's spectrum- and mutation-based features applicable to
        detect faulty statements.
\end{itemize}

%reviewer comments: {\color{red}{The paper should include a discussion of more recent work on neural net based fault localization: "Neural attribution for semantic bug localization in student programs" (NeurIPS 2019).
}}

As in  FL work~\cite{b2016learning,DeepFL,TraPT} using
Defects4J, we used the setting of leave-one-out cross validation on
the faults for each individual project (i.e., within-project setting).
Specifically, we use one bug (i.e., with one buggy statement or
method) as testing, and the remaining bugs in a project for training.
%
%there can be $n$ faulty versions for a project and {\color{red}{each
%    faulty version is a bug}}. We divide $n$ faulty versions into two
%groups: $n-1$ versions for training our {\tool} and one version for
%testing.
%
%This setting provides sufficient training data for the models because
%a project contains a large number of buggy statements.
%We performed the cross-project setting in RQ4.

%
%For example, the project Closure in Defects4J has 100,000+ training
%instances~\cite{li2019deepfl}.

%Each buggy version can contain a very large
%number of training instances, i.e., statements executed by failed
%tests. For example, the project Closure in Defects4J already has
%100,000+ training instances~\cite{li2019deepfl}.

%We separate them by one bug as the test data to get the result and other $-1$ bugs as training data to train the whole model.
%We do the experiment like this because there are many existing approaches \cite{Exisiting} also do this like.

\textit{Tuning {\tool} and the baselines.}
%We tuned approaches to get the best performance of our model.
%For simplicity, we use the same set of parameter settings for approaches in both of the
%above mentioned experimental settings once the best settings are identified.
%Like any other deep learning based approaches,
We tuned our model with the following key hyper-parameters to obtain
the best performance: (1) Epoch size (i.e., 100, 200, 300); (2) Batch
size (i.e., 64, 128, 256); (3) Learning rate (i.e., 0.001, 0.003,
0.005, 0.010); (4) Vector length of word representation and its output
(i.e., 150, 200, 250, 300); (5) Convolutional core size (i.e.,
$3\times3,\:5\times5,\:7\times7,\:9\times9,$ and $\:11\times11$); (6)
The number of convolutional core (3, 5, 7, 9, and 11).

As for {\em word2vec}, for a method, we consider all tokens in the
source code order as a sentence.
%By considering all sentences collected from training dataset, we build
%our token representation list to use in \tool.
We tune the following hyper-parameters for DeepFL (using only the
features relevant to statements): Epoch number (5, 10, 15, ..., 60),
Loss Functions (softmax, pairwise), and learning rate (0.001, 0.005,
0.010).

%reviewer comments: {\color{red}{Many experimental details are missing. To give a few examples, when word2vec methods are used, how are the vocabularies constructed, what their sizes are, which representations are trained on which datasets, for how long, and so on.
%}}

\vspace{3pt}
\noindent\textbf{RQ2: Method-level Fault Localization Comparison.}

%In this work, for fault localization, we also designed our model in method level in order to work as some other existing fault localization approaches. We aim to compare our approach in different levels in order to prove that we model can work well on all levels.

\textit{Baselines:} We also compare our approach with the following
state-of-the-art approaches that localize faulty methods.

% We compare our code representation with the following state-of-the-art fault localization approaches at method-level:

\textbf{MULTRIC}~\cite{MULTRIC} is a learning-based approach to
  combine different spectrum-based ranking techniques using
  learning-to-rank for effective fault localization.

\textbf{FLUCCS}~\cite{FLUCCS} is a learn-to-rank based technique
using spectrum-based scores and change metrics (e.g., code
churn and complexity metrics) to rank program elements.
%extended from SBFL with code and changed metrics that have been studied in the context of defect prediction, such as size, age and code churn.

\textbf{TraPT}~\cite{TraPT} is a learn-to-rank technique to
  combine spectrum-based and mutation-based fault localization.
%  techniques for elective fault localization.

\textbf{DeepFL}~\cite{DeepFL} is a DL-based model
  to learn the existing/latent features
  from multiple aspects of test cases and a program. We used all the
  features of DeepFL in this method-level study.

  %for precise fault
  %  localization. We use the whole approach in this RQ.

\textit{Tuning {\tool} and the baselines.}  Similar to RQ1, we perform
our experiments using leave-one-out cross validation on the faults for
each project. We use the same settings in RQ1 to train our model.
Note that in DeepFL paper~\cite{DeepFL}, {\bf DeepFL},
\textbf{MULTRIC}, \textbf{FLUCCS}, and \textbf{TraPT} have been
evaluated using leave-one-out cross validation and other settings on
the same data set of Defects4J V1.2.0.
%
%In \cite{li2019deepfl}, DeepFL significantly outperformed the above
%three learning-to-rank tools. Therefore, we only ran and tuned DeepFL
%to verify the results published in \cite{li2019deepfl}.
{\tool} is also evaluated on Defects4J V1.2.0 using the same settings
and procedure as DeepFL. Thus, we used the result on the numbers of
detected bugs reported in DeepFL~\cite{DeepFL} for those models.
%
%we directly reuse the results for those tools
%reported in DeepFL paper~\cite{li2019deepfl}.

%of \textbf{MULTRIC}, \textbf{FLUCCS}, and \textbf{TraPT} reported in
%\cite{li2019deepfl}, and run DeepFL with its published best settings.

%As for baselines, we directly bring the results from existing approach DeepFL \cite{DeepFL} for most baselines results. As for the rest baselines, they do not need to do the parameter tuning. So we directly use their model to get the results.
%\noindent\textbf{RQ3. Impact Analysis of Different Matrix Enhancing Techniques.} %Error-exhibiting Lines, Test Case Re-ordering, and Statements Dependency.}
%How do different matrix enhancing techniques including error-exhibiting lines, test case re-ordering, and statements dependency affect the overall performance of {\tool}?

%\noindent\textbf{RQ4. Impact Analysis of Different Representations.} How do different types of representation affect the overall performance of {\tool}?
%How do different types of representation for the methods/statements affect the overall performance of {\tool}?

	\vspace{3pt}

\noindent\textbf{RQ3: Impact Analysis of Different Matrix Enhancing
  Techniques.}
%We aim to evaluate the impact of our CC matrix enhancing techniques on
%performance.
We evaluate the impact of the following techniques on accuracy: (1)
test case ordering algorithm utilizing the EE lines ({\bf Order}); (2)
 statements' dependencies
%modeled by using execution paths of statements and data-flow graph
({\bf StatDep}). We first build a base model by using only the
spectrum- and mutation- based matrices in {\tool} (without using the
above techniques), then apply the above techniques on the matrices to
build two variants of {\tool}: \{\emph{Base+Order}\}, and
\{\emph{Base+Order+StateDep} ({\tool})\}. We train each variant
using the same settings as in RQ1.  Due to space limit, we show only
the analysis results obtained in the within-project setting for
method-level FL.
		
\vspace{3pt}
\noindent\textbf{RQ4: Impact Analysis of Learning Representations.}
%We used the matrix enhancing techniques to generate new
%representations for spectrum-based and mutation-based matrices, then
%we conducted representation learning to learn a vector to represent a
%statement/method.
We have the following representation learning schemes: the enhanced
spectrum-based CC matrix ({\bf NewSpecMatrix}) and the enhanced
mutation-based CC matrix ({\bf NewMutMatrix}). We also have source
code representation ({\bf CodeRep}) and textual similarity between
source code and error messages in failing tests ({\bf TextSim}). To
test the impact of those representation learning schemes on accuracy,
we built a base model using only {\em NewSpecMatrix}, and three other
variants: \{\emph{NewSpecMatrix+NewMutMatrix}\},
\{\emph{NewSpecMatrix+ NewMutMatrix+CodeRep}\}, and
\{\emph{NewSpecMatrix+ NewMutMatrix+CodeRep+TextSim}\}. We trained
each variant using the same settings as in RQ1. Due to space limit, we
show only the results for the within-project setting for method-level
FL.

		%We perform an analysis to test the impact of components.
		%Due to space limit, we show only effectiveness analysis on components results in the {\bf within-project setting at method-level localization.}

%To perform sensitivity analysis, we add each element into the model one by one in each of the two settings with different parameters.

\iffalse
\vspace{5pt}
\noindent\textbf{RQ4-Analysis Approach (Advanced Models Analysis).}
We use a basic CNN on top of the vectors learned from code coverage matrices and code representation learning. We replace the basic CNN with advanced models to see whether the performance can be improved or not.
%The advanced model we want to try to replace the basic CNN model in our approach include:
We test two recent CNN-based models:
\begin{itemize}
	
	\item \textbf{H-CNN}~\cite{HCNN} is a combination model of self-attention mechanisms, convolutional filters and a hierarchical structure to do the classification which is both highly accurate and fast to train.
	
	\item \textbf{Densely CNN} (D-CNN)~\cite{DCNN} connects each layer to every other layer in a feed-forward fashion. Whereas traditional convolutional networks with L layers have L connections - one between each layer and its subsequent layer - its network has L(L+1)/2 direct connections.
	
\end{itemize}

%In this RQ, we try to replace all CNN in our model with the advanced model listed above in order to evaluate if the advanced model can help improve the performance of the whole approach.
\fi

\vspace{3pt}
\noindent\textbf{RQ5: Cross-project Analysis.}  We also setup
the cross-project scenario: testing one bug in
a project, but training a model on all of the bugs of other
projects. For a project, we test every bug and sum up the total
number of bugs in the project that can be localized in the
cross-project scenario.
%Due to page limit, we only show the results of method-level fault
%localization for the cross project.

\vspace{3pt}
\noindent\textbf{RQ6: {\tool's} Fault Localization Performance on C
  Code.} We also evaluated {\tool} on C projects from the benchmark
dataset, ManyBugs~\cite{manybugs,LeGoues15tse}, with 185 bugs from 9
projects. We used the same model in RQ1 for statement-level FL
and the model in RQ2 for method-level FL.

%\vspace{3pt}
%\noindent\textbf{RQ7: Sensitivity Analysis.} We do the sensitivity analysis on \tool for textual similarity between source code and error messages in failing
%tests ({\bf TextSim}) and (will add one more component here).

\subsection{Experimental Results}

\subsubsection{RQ1-Results ({\bf Statement-level Fault Localization Comparison})}\label{rq1-results}
\begin{table}[t]
	%\vspace{-10pt}
	\caption{RQ1. Results of comparative study for statement-level fault localization. P\% = $|$Top-1$|$/\{395 Bugs\}}
	\vspace{-5pt}
\tabcolsep 4.5pt
	{\small
	\begin{center}
		\renewcommand{\arraystretch}{1}
		\begin{tabular}{l|p{0.8cm}<{\centering}p{0.8cm}<{\centering}p{0.8cm}<{\centering}p{0.8cm}<{\centering}p{0.8cm}<{\centering}p{0.8cm}<{\centering}}
			\hline
			Approach & Top-1 & Top-3 & Top-5 & P\% & MFR & MAR \\
			\hline
			
			Ochiai & 17& 88 & 115 & 4.3\% & 54.29 & 71.32 \\
			Dstar & 19 & 92 & 115 & 4.8\% & 48.67 & 69.51 \\
			MUSE & 26 & 47 & 63 & 6.6\% & 36.34 & 48.73 \\
			Metallaxis & 24& 81 & 108 & 6.1\% & 34.59 & 49.21\\
			RBF & 12 & 37 & 52 & 3.0\% & 22.54 & 57.47 \\
			DeepFL& 39 & 114 & 129 & 9.9\% & 24.09 & 31.28 \\\hline
			{\bf \tool}	& \textbf{71} & \textbf{128} & \textbf{142} & \textbf{18.0\%} & \textbf{20.32} & \textbf{28.63} \\
			\hline
		\end{tabular}
		
		\label{fig:statement_level}
	\end{center}
    }
	\vspace{-10pt}
\end{table}

As seen in Table~\ref{fig:statement_level}, {\tool} improves over
the state-of-the-art statement-level FL baselines.
%It improves all of the baselines at statement-level in every metric. 
Specifically, {\tool} improves Recall at Top-1 by 317.6\%, 273.7\%,
173.1\%, 195.8\%, 491.7\%, and 82.1\% in comparison  with Ochiai,
Dstar, Muse, Metallaxis, RBF, and DeepFL.

%Reviewer Comments: {\color{red}{As for the metrics, besides TOP-1/MAR/MFR,  I think it’s also necessary to show the TOP-3/TOP-5 results to show a more comprehensive comparison.
%}}

%will add comparision on other metrics after putting all results into the table.

%\textbf{In-depth Case Studies.}

%\subsection{A case study}

We examined the results and report the following. The key reason for
the spectrum-based FL approaches fail to localize the buggy statements
is that they give the same suspiciousness score to the statements at
the same nested level. For the mutation-based FL approaches, the key
reason for not being able to localize the buggy statements/methods is
that the fix requires a more sophisticated change than a mutation. Let
us take an example. In Fig.~\ref{fig:discu_1}, the fault is caused
by an incorrect variable.  To fix it, the variable was changed from
\textit{pos} to \textit{pt} at line 14. The state-of-art
spectrum-based approaches cannot localize this fault because lines 6,
7, 13, and 14 have the same score (They were executed in both passing
and failing test cases).
For the mutation-based FL approaches, there is none of mutation
operators that changes the variable $pos$ into $pt$ in a method call
at the buggy line 14. Thus, they cannot observe the impact of
mutations on the code coverage. As a consequence, they cannot locate
the buggy line 14.

%For the mutation-based approaches, no mutator can fix this unique
%fault.
%Thus, the mutation based approaches also cannot deal with this fault.

\begin{figure}[t]
	\vspace{-26pt}
	\centering
	\lstset{
%		numbers=left,
%		numberstyle= \tiny,
%		keywordstyle= \color{blue!70},
%		commentstyle= \color{red!50!green!50!blue!50},
%		frame=shadowbox,
%		rulesepcolor= \color{red!20!green!20!blue!20} ,
%		xleftmargin=7em,xrightmargin=7em, aboveskip=1em,
%		framexleftmargin=1em,
%		language=Java,
%		basicstyle=\scriptsize\ttfamily,
          %		escapeinside= {(*@}{@*)}
       		numbers=left,
		numberstyle= \tiny,
		keywordstyle= \color{blue!70},
		commentstyle= \color{red!50!green!50!blue!50},
		frame=shadowbox,
		rulesepcolor= \color{red!20!green!20!blue!20},
                xleftmargin=1.5em,xrightmargin=0em, aboveskip=1em,
		framexleftmargin=1.5em,
                 numbersep= 4pt,
		language=Java,
                basicstyle=\scriptsize\ttfamily,
                numberstyle=\scriptsize\ttfamily,
                emphstyle=\bfseries,
		%basicstyle=\tiny\ttfamily,
		escapeinside= {(*@}{@*)}
	}
	\begin{lstlisting}[title = \quad]
public void translate(CharSequence input, Writer out)...{
  ...
  int pos = 0;
  int len = input.length();
  while (pos < len) {
    int consumed = translate(input, pos, out);
    if (consumed == 0) {
      char[] c=Character.toChars(Char...codePointAt(...));
      out.write(c);
      pos+= c.length;
      continue;
    }
    for (int pt = 0; pt < consumed; pt++) {
(*@\scriptsize \color{red}{- \quad pos += Char.charCount(Char.codePointAt(input,pos));}@*)
(*@\scriptsize \color{cyan}{+ \quad pos += Char.charCount(Char.codePointAt(input, pt));}@*)
   }
 }
}
\end{lstlisting}
\vspace{-5pt}
\caption{An Example from Defects4J}
\label{fig:discu_1}
\vspace{-5pt}
\end{figure}

\begin{figure}[t]
	\centering
	\includegraphics[width=2in]{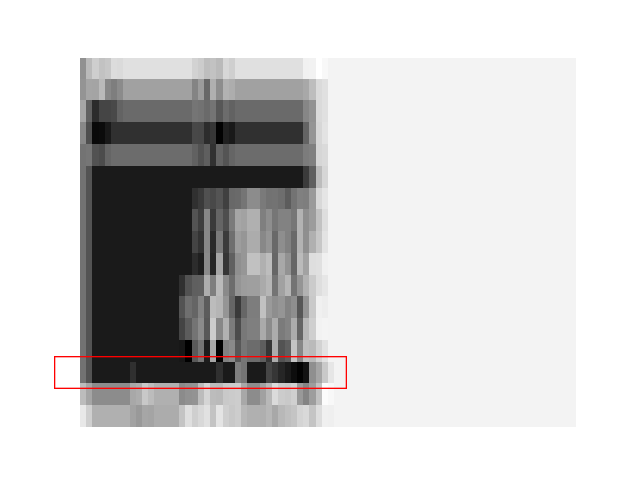}
        \vspace{-10pt}
	\caption{A Feature Map Produced by CNN for Fig.~\ref{fig:discu_1}}
	\label{fmap} 
	\vspace{-10pt}
\end{figure}

To gain insights, we performed a visualization of a feature
map for this case.
%This is a common technique in image processing with CNN. Note that,
During training, CNN learns the values for small windows,
called {\em filters}. The feature maps of a CNN capture the result of
applying the filters to an input matrix. That is, at each layer, the
feature map is the output of that layer. In image processing,
visualizing a feature map for an input helps gain understanding on
whether the model detects some part of our desired object and what
features the CNN observes. Fig.~\ref{fmap} shows a feature map for the
example in Fig.~\ref{fig:discu_1}. We can see that around the lines
6--8 and 13--14, the feature map is visually dark. Without ordering
(i.e., a random order of test cases), the feature map does not exhibit
such visualization.

To further study the impacts of the ordering and data dependencies, we
modified {\tool} in the following settings: 1) {\em No ordering + No
dependencies}: the buggy line 14 is ranked at 43th; 2) {\em No
ordering + dependencies}: it is ranked at 29th; 3) {\em Ordering + No
dependencies}: it is ranked at 7th; and 4) {\em Ordering + dependencies}: it
is ranked at the top.

\subsubsection{RQ2-Results ({\bf Method-level Fault Localization Comparison})}

\begin{table}[t]
	\caption{RQ2. Results of Comparative Study for Method-level Fault Localization. P\% = $|$Top-1$|$/\{395 Bugs\}}
	\vspace{-5pt}
        \tabcolsep 2pt
	{\small
	\begin{center}
		\renewcommand{\arraystretch}{1}
		\begin{tabular}{l|p{1cm}<{\centering}p{1cm}<{\centering}p{1cm}<{\centering}p{1cm}<{\centering}p{1cm}<{\centering}p{1cm}<{\centering}}
			\hline
			Approach & Top-1 & Top-3 & Top-5 & P\% & MFR & MAR \\
			\hline
			
			MULTRIC & 80& 163& 195 & 20.3\% & 37.71 & 43.68 \\
			FLUCCS & 160& 249 & 275 & 40.5\% & 16.53 & 21.53 \\
			TraPT & 156& 249 & 281 & 39.5\% & 9.94 & 12.70 \\
			DeepFL & 213 & 282 & 305 & 53.9\% & 6.63 & \textbf{8.27} \\\hline
			{\bf {\tool}} & \textbf{245}& \textbf{294} & \textbf{311} & \textbf{62.0\%} & \textbf{5.94} & 8.57 \\
			\hline
		\end{tabular}
		\label{fig:method_level}
		
	\end{center}
}
%\vspace{-10pt}
\end{table}

As seen in Table~\ref{fig:method_level}, {\tool} improves~Recall at
Top-1 by 206.3\%, 53.1\%, 57.1\%, and 15.0\% over MULTRIC, FLUCCS,
TraPT, and DeepFL, respectively. {\tool}'s MAR is slightly higher than
DeepFL's (3.6\% higher). On average, {\tool} ranks the correct
elements higher than DeepFL, as its MFR is lower (10.4\% lower).

%Ranking correct elements higher saves effort from developers.

%Reviewer Comments: {\color{red}{As for the metrics, besides TOP-1/MAR/MFR,  I think it’s also necessary to show the TOP-3/TOP-5 results to show a more comprehensive comparison.
%}}

\begin{figure}[t]
\centering
\includegraphics[width = 3.6in]{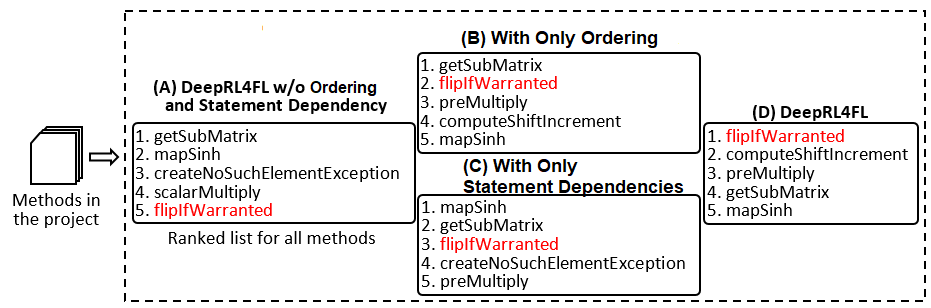}
\vspace{-6pt} %-12pt
\caption{Ordering and Statement Dependencies Affect Ranking}
\vspace{-5pt}
\label{rq2ex} 
\end{figure}

%To further examine our results, we report the following.  MULTRIC,
%FLUCCS, and TraPT locates the fewer bugs, as they did not fully
%utilize different types of information for a method so that they
%missed some bugs.  MULTRIC only relies on the spectrum-based
%scores. FLUCCS utilizes spectrum-based features and the code/change
%metrics.  TraPT derives the ranking features from the mutation-based
%testing tools and the analysis of test case code with test
%outputs/messages.  DeepFL can locate more bugs than the above three
%learning-to-rank based approaches, as it utilizes four types of
%information: (1) scores from spectrum-based testing tools, (2) scores
%from mutation-based testing tools; (3) code/change metrics; and (4)
%the textual similarities between failed tests and source code.

%Through an overlap analysis between DeepFL and {\tool}, we summarized
%two main reasons why we can locate more bugs than DeepFL: (1) {\tool}
%does not directly use the scores from the spectrum and mutation-based
%formulas for learning. {\tool} directly enriches the matrices with
%test case relations: ordering the test cases based on code coverage
%results.  (2) {\tool} also enriches the code coverage matrices with
%the statement dependences of a method. The above matrix enrichment
%enables the later learning models to capture more characteristics of a
%method.

The spectrum-based and mutation-based FL approaches fall short of
DeepFL and {\tool}. A key reason is that they consider only dynamic
information in test cases, while DeepFL and our model use both static
and dynamic information. In comparison with DeepFL, we further
analyzed the bugs that our tool can locate, but DeepFL missed. We
found that the mean first rank of a buggy method in the ranking lists
of potential buggy methods returned by DeepFL is 7.08. Without the
ordering and statement dependency in our model, the mean first rank is
6.84. With only ordering in our model, the mean first rank is
2.82. With only dependency in our model, the mean first rank is
4.45. With both ordering and dependency, our model can locate the bugs
that DeepFL missed.

Let us use an example in Defects4J (Fig.~\ref{rq2ex}) that our model
detected but DeepFL missed. The (buggy) method \code{flipIfWarranted}
together with the other methods in the project were fed into four
variants of our model. As seen, with the setting in which both
ordering and statement dependencies are removed,
\code{flipIfWarranted} is ranked 5th in the list of all methods. For
the setting with only ordering, it is ranked at 2nd place. For the
setting with only statement dependencies, it is ranked 3rd. With both,
our model ranks the buggy method \code{flipIfWarranted} at the 1st
position. This analysis shows that ordering test cases and statement
dependencies are the key drivers that help our model locate more bugs
than DeepFL.

		\subsubsection{RQ3-Results ({\bf Impact Analysis of Different Matrix Enhancing Techniques})}
		\begin{table}[t]
			\caption{RQ3. Ordering (Order) and Adding Dependencies (StateDep) in Method-level FL. P\% = $|$Top-1$|$/\{395 Bugs\}}
			\vspace{-10pt}
                        \tabcolsep 2pt
			{\scriptsize
				\begin{center}
					\tabcolsep 3pt
					\renewcommand{\arraystretch}{1}
					\begin{tabular}{l|p{0.8cm}<{\centering}p{0.8cm}<{\centering}p{0.6cm}<{\centering}p{0.6cm}<{\centering}}
						\hline
						Variants & Top1 & P\% & MFR & MAR \\
						\hline
					    Base ({\tool} $w/o$ Order,StateDep) & 173 & 43.8\% & 8.23 & 10.27 \\
					%	Base + Errorlines & &&&  &  &  \\
					%	Base + ReOrder & &&&  &  &  \\
						Base + Order & 226 & 57.2\% & 6.57 & 8.97 \\
						Base + Order + StateDep ({\bf {\tool}}) & 245 & 62.0\% & 5.94 & 8.57 \\
						%\tool$w/o$ TCEE & 219 &&& 55.4\% & 6.72 & 9.19 \\
						%\tool$w/o$ SBM & 199&& &50.4\% & 7.36 & 9.41 \\
						%\tool$w/o$ MBM & 227&& & 57.5\% & 6.69 & 8.73 \\
						%\tool$w/o$ CRL (=CoRL+SD) & 232 &&& 58.7\% & 6.13 & 8.69 \\
						%\tool$w/o$ SD (=CoRL+CRL) & 226 &&& 57.2\% & 6.57 & 8.97 \\
						%\tool$w/o$ Sim & 221 &&& 55.9\% & 6.84 & 9.18 \\
						\hline
					\end{tabular}
					\label{fig:sensitivity_2}
				\end{center}
			}
			\vspace{-10pt}
		\end{table}

		%In this RQ, we conduct ablation analysis to study the impact of the following components of our \tool: 	(1) spectrum based improved code coverage matrix (SBM);
		%(2) mutation based improved code coverage matrix (MBM);
		%(3) code representation representation (CRL);
		%(4) execution paths of statements and data-flow graph for modeling relations among statements (EP-DF);
		%and (5) textual similarity between source code and error messages in failing tests (Sim).
		
		%As seen in Table~\ref{fig:sensitivity_2}, we evaluate the error-exhibiting lines {\bf{EL}}, test case re-ordering, and statements dependency by taking \tool with error-exhibiting lines as basic model. It can localized XX bugs for top1 and have YY\% and YY\% on MFR and MAR. By adding test case re-ordering {\bf{TCR}}, the top1, MFR, and MAR have an relevant improvement by XX, XX, XX. And by adding statements dependency {\bf{SD}}, model can fix XX bugs on top1 and can improve the MFR and MAR by XX and XX.

%	Base + ReOrder & 226 &&& 57.2\% & 6.57 & 8.97 \\
%Base + ReOrder + StateDep ({\tool}) & 245 &&& 62.0\% & 5.94 & 8.57 \\

\begin{figure}[t]
	\centering
	\includegraphics[width=3.3in]{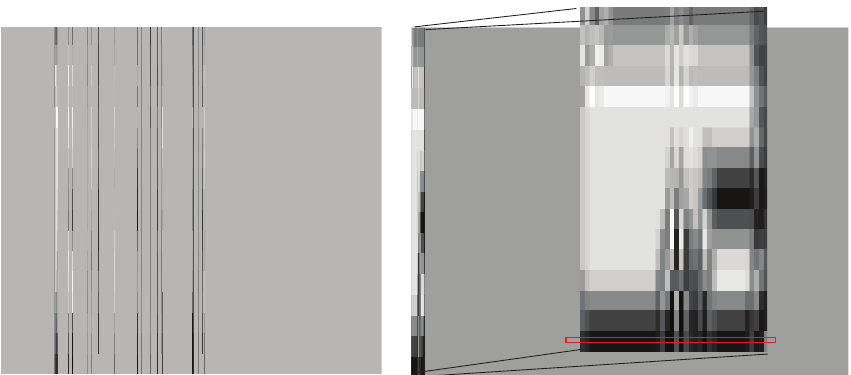}
	\caption{Visually Darker Lines around Buggy Statement}
	\label{fmap2}
	\vspace{-6pt}
\end{figure}

%Tien
Table~\ref{fig:sensitivity_2} shows that our matrix
enhancing techniques positively contribute to {\tool}. Specifically,
comparing \{\emph{Base}\} with \{\emph{Base+Order}\}, ordering the test cases
can improve every metric. \emph{Order} helps localize 53 more bugs (13.4\%)
using Top-1. It helps improve MFR and MAR by 20.1\% and 12.7\%,
respectively, showing that ordering can help {\tool} push the faulty
methods higher in the ranked list.

Comparing \{\emph{Base+Order}\} with \{\emph{Base+Order+StateDep}\}, we
see that modeling dependencies into matrices is useful~to improve the
performance of {\tool}. StateDep can improve 8.4\%, 9.6\%, and 4.5\% in
Top-1, MFR, and MAR.

% --- Tien
To further study {\em the impact of the ordering}, we visualize the
feature maps for the 53 bugs that {\em Order} can detect and {\em
Base} did not. Those are the cases where ordering helps.  Visualizing
the feature maps for those inputs allows us to understand what
features the CNN detects in both cases of ordering and
no-ordering. Moreover, that also allows us to see if ordering can help
the CNN model learns better the discriminative features~in locating
the buggy statements.
To do so, for each of those bugs, we used the CNN model as part of
{\em Base} and {\em Order} to produce two feature maps: one
corresponds to {\em Base} (no~ordering) and one to {\em Order}. We
then visualized and compared those feature maps as
gray-scale images. The CNN model generates 9 feature maps as the
output from 9 convolutional cores.

In all the bugs, we observe the same phenomenon. Let~us take an
example. Fig.~\ref{fmap2} shows two feature maps for one of those
bugs. The left image is for {\em Base} (without ordering), and the right
one is for {\em Order} (with ordering).
%ones on the left and on the right are for {\em
%Base} (without ordering) and for {\em Order} (with ordering),
%respectively.
We zoom out the leftmost columns in the right feature map. The row
corresponding to the buggy line is in the red rectangle.  With
ordering, one of those 9 feature maps has visually darker lines around
the buggy statement. In contrast, without ordering, all the feature
maps are similar to the one on the left, i.e., do not show any clear
visual lines. In brief, with ordering, the CNN model, which focuses on
the relations of neighboring cells, can detect the features along the
buggy~statement.

\subsubsection{RQ4-Results ({\bf Impact Analysis of Learning Representations})}

		\begin{table}[t]
			\caption{RQ4. Results of Learning Representations in Method-level FL. P\% = $|$Top-1$|$/\{395 Bugs\}}
			\vspace{-10pt}
			{\scriptsize
				\begin{center}
					\tabcolsep 3pt
					\renewcommand{\arraystretch}{1}
					\begin{tabular}{l|p{0.6cm}<{\centering}p{0.5cm}<{\centering}p{0.4cm}<{\centering}p{0.4cm}<{\centering}}
						\hline
						Variants & Top-1 & P\% & MFR & MAR \\
						\hline
						
						Base (NewSpecMatrix)& 189 & 47.8\% & 8.09 & 9.91 \\
						Base+NewMutMatrix &  212 & 53.7\% & 7.43 &9.56 \\
						Base+NewMutMatrix+CodeRep &  221 & 55.9\% & 6.84 & 9.18 \\
						Base+NewMutMatrix+CodeRep+TextSim ({\bf {\tool}})& 245 & 62.0\% & 5.94 & 8.57 \\
						
						%SBM & &&&  & &  \\
						%\tool$w/o$ TCEE & 219 &&& 55.4\% & 6.72 & 9.19 \\
					%	SBM+MBM &  232 &&& 58.7\% & 6.13 & 8.69 \\
					%	SBM+MBM+CRL(All)& 245 &&& 62.0\% & 5.94 & 8.57 \\
						%\tool$w/o$ CRL (=CoRL+SD) & 232 &&& 58.7\% & 6.13 & 8.69 \\
						%\tool$w/o$ SD (=CoRL+CRL) & 226 &&& 57.2\% & 6.57 & 8.97 \\
						%\tool$w/o$ Sim & 221 &&& 55.9\% & 6.84 & 9.18 \\
						\hline
					\end{tabular}
					\label{fig:sensitivity_1_1}
				\end{center}
			}
			\vspace{-5pt}
		\end{table}
	
	%Base (NewSpecMatrix)& &&&  & &  \\
	%Base + NewMutMatrix &  232 &&& 58.7\% & 6.13 & 8.69 \\
	%Base + NewMutMatrix + CodeRep &  232 &&& 58.7\% & 6.13 & 8.69 \\
	%Base + NewMutMatrix + CodeRep+TextSim ({\tool})& 245 &&& 62.0\% & 5.94 & 8.57 \\
	
Table~\ref{fig:sensitivity_1_1} shows that our representation
learning~has positive contributions. Comparing \{\emph{Base}\}
with \{\emph{Base+ NewMutMatrix}\}, we can see that
%
%the representation learning on
mutation-based matrices can help locate 23 more bugs using Top-1 and
improve MFR and MAR by 8.2\% and 3.5\%. By adding code
representation learning,
%into the model,
we improve {\tool} to localize 9 more bugs and gain an increase on
MFR and MAR by 7.9\% and 4.0\%, respectively. Furthermore, TextSim
also positively contributes to our model.
For statement-level FL, code representation is also useful, improving
Top-1 from 65 to 71 bugs, i.e., 9.2\% (not shown).

\subsubsection{RQ5-Results ({\bf Cross-project Analysis})}

\begin{table}[ht]
	\caption{RQ5. Cross-project versus Within-project Results}
	\vspace{-10pt}
\tabcolsep 3.5pt
	{\footnotesize
		\begin{center}
			\begin{tabular}{p{0.9cm}<{\centering}|p{0.7cm}<{\centering}p{0.7cm}<{\centering}p{0.5cm}<{\centering}p{0.7cm}<{\centering}|p{0.7cm}<{\centering}p{0.7cm}<{\centering}p{0.5cm}<{\centering}p{0.7cm}<{\centering}}\hline \cline{2-9}	
				\multirow{2}{*}{Projects}& \multicolumn{4}{c|}{Cross-project} & \multicolumn{4}{c}{Within-project}\\\cline{2-9}
				& Top-1&P\%&MFR&MAR& Top-1&P\%&MFR&MAR\\\hline
				Chart   & 13 & 50.0\% & 3.15 & 5.62 &    15 & 57.7\% & 2.85 & 4.65  \\ 
				Time    & 13 & 48.1\% & 9.78 & 14.70 &   14 & 51.9\% & 8.41 & 13.33   \\ 
				Math    & 61 & 57.5\% & 3.81 & 4.88 &    64 & 60.4\% & 2.93 & 4.83   \\ 
				Closure & 71 & 53.4\% & 11.70 & 15.23 &  73 & 54.9\% & 9.38 & 12.37  \\ 
				Mockito & 12 & 31.6\% & 11.42 & 16.42 &  14 & 36.8\% & 9.39 & 15.11  \\
				Lang    & 47 & 72.3\% & 2.13  & 2.49  &  50 & 76.9\% & 1.97 & 2.31\\
				\hline
				%\vspace{1pt}
			\end{tabular}
			\label{RQ4}
		\end{center}
	}
	\vspace{-10pt}
\end{table}

As seen in Table~\ref{RQ4}, {\tool} achieves better results in the
within-project setting than in the cross-project one. This is
expected as the training and testing data is from the same project in
the within-project setting, thus a model may see similar faults.

In the cross-project setting, {\tool} correctly detects 217 bugs at
Top-1 in comparison with the best result (207 bugs) from the
baselines. In the within-project setting, {\tool} correctly
detects 230 bugs at Top-1 in comparison with 80/160/156/213 bugs
(not shown) from the baseline models
MULTRIC/FLUCCS/TraPT/DeepFL, respectively.

%Compared with other method-level baselines, {\tool}'s cross-project
%results (i.e., 217 detected bugs for Top-1) is higher the
%within-project results of all other studied method-level baselines,
%MULTRIC/FLUCCS/TraPT/DeepFL (i.e., 80/160/156/213 for Top-1). The
%cross-project result of DeepFL is 207 for Top-1, and {\tool} can
%localize 20 more bugs.

\vspace{0.02in}
\noindent {\bf Time Complexity.} On average, training time is 350-380 minutes per project in the
cross-project setting, and 120-130 minutes per project in the
within-project setting. Once the model is trained, the prediction time
per fault is 2-7 seconds in both the cross- and within-project
settings.

%{\color{red}{As for the metrics, besides TOP-1/MAR/MFR,  I think it’s also necessary to show the TOP-3/TOP-5 results to show a more comprehensive comparison.
%}}

\subsubsection{RQ6-Results ({\bf Performance on C Code})}

As seen in Table~\ref{RQ5}, {\tool} can localize 27 faulty
statements and 98 faulty methods with only Top-1 statements and
methods. The empirical results show that the performance of
{\tool} on the C projects is consistent with the one on the Java
projects. Specifically, at the statement level, the percentages of the
total C and Java bugs that can be localized are similar, i.e., 14.6\%
vs. 18.0\%, respectively. At the method level, the percentages of the
total C and Java bugs that can be localized are also consistent, i.e.,
53.0\% vs. 62.0\%, respectively.

\begin{table}[t]
	\caption{RQ6. ManyBugs (C Projects) versus Defects4J (Java Projects). P\% = $|$Top-1$|$/\{Total~Bugs~in Datasets\}}
	\vspace{-5pt}
	{\small
		\begin{center}
                        \tabcolsep 2.7pt
			\begin{tabular}{p{1.3cm}<{\centering}|p{0.8cm}<{\centering}p{0.8cm}<{\centering}p{0.6cm}<{\centering}p{0.8cm}<{\centering}|p{0.8cm}<{\centering}p{0.8cm}<{\centering}p{0.6cm}<{\centering}p{0.8cm}<{\centering}}\hline \cline{2-9}	
				\multirow{2}{*}{Level}& \multicolumn{4}{c|}{ManyBugs (C projects)} & \multicolumn{4}{c}{Defects4J (Java projects)}\\\cline{2-9}
				        & Top-1  & P\%    & MFR   & MAR   & Top-1 & P\%     & MFR & MAR \\
				\hline
				Statement   & 27 & 14.6\% & 25.74 & 31.33 & 71 &  18.0\% & 20.32 & 28.63 \\ 
				Method   & 98 & 53.0\% & 6.91  & 9.89  & 245 & 62.0\% & 5.94  & 8.57  \\  
				\hline
				%\vspace{1pt}
			\end{tabular}
			\label{RQ5}
		\end{center}
	}
	\vspace{-10pt}
\end{table}

%{\color{red}{As for the metrics, besides TOP-1/MAR/MFR,  I think it’s also necessary to show the TOP-3/TOP-5 results to show a more comprehensive comparison.
%}}

\subsubsection{Threats to Validity}
%We have identified the following threats to the validity:

\textbf{i) Baseline implementation.}~For comparative study,
we implemented Ochiai, Dstar, MUSE, Metallaxis, and RBF-neural-network
for statement-level FL. We followed the paper~\cite{DeepFL} to
implement MUSE and Metallaxis using PIT-1.1.5.  RBF-neural-network
approach is built for artificial faults and our real bug dataset
cannot match the requirements.
\textbf{ii) Result generalization.} Our comparisons with the
baselines were only carried out on the Defects4J dataset.
%, which is a widely used benchmark for FL research.
Further evaluation on other datasets should be done.

%\noindent\textbf{Selection of programming languages.} In this study,
%we only apply our approach on Java code. Thus, we cannot claim that
%our approach is generic for all programming languages. However, our
%key ideas that drives {\tool} to outperform others baselines are
%general across programming languages: modeling relations between test
%cases and modeling statements' dependencies.

\subsubsection{Limitations}

%In this section, we would like to use real code example to make a discussion about our approach performance.

%\input{sections/discu_1}
%\\textbf{Limitations}:
%\subsection{Limitations}

The quality of test cases is important for our approach. If there are
only a couple of passing test cases or the crash occurs far apart from
the faulty method, {\tool} does not learn a useful representation
matrix to localize the faults. It does not work well on locating the
faults that require statement additions to fix (all of the baselines
in this paper do not either). Moreover, it does not work well for
short methods, as they provide less statement dependencies. It is also
hard for our model to localize the uncommon faults. Because it is
DL-based, if there is a very uncommon fault that may not be
seen in the training dataset, it will not work correctly.

%\input{sections/discu_2}

%{\bf Limitations of our approach.}
%First, \tool do not work well on locating the faults that require statement additions to fix as well as all of the studied baselines in this paper. 
%Because if there needs to add some statements to fix the bug, 
%we cannot let our model analysis the relationship between original source code and added statements which means our model cannot deal with the problem of locating the correct position for adding statements well. 
%Second, it is hard for \tool to locate the faults that happens in very short methods. 
%Because our approach needs to do analysis on the relationship in order to locate the faults, very short method cannot provide enough information to make our approach work well.
%Second, it is hard for \tool to locate the faults that have 

%error massages reported the error line in other statements within other method which has no close relationship. 
%Because our approach would like to use execution path and data flow graph to catch the relationship between statements, if the relationship is far away (i.e. The relationship pass through several statements and then link them together), it is hard for our approach to detect this relationship and use it to locate the faults.

%\input{sections/illustration}

\section{Related Work}

\textbf{Fault Localization (FL).}  The Spectrum-based Fault Localization
(SBFL), e.g.,~\cite{Ochiai,abreu2007accuracy,
  jones2005empirical,keller2017critical, liblit2005scalable,
  lucia2014extended,naish2011model, wong2007effective,
  zhang2011localizing}, has been intensively studied in the
literature.  Tarantula \cite{jones2001visualization}, SBI
\cite{liblit2005scalable}, Ochiai \cite{Ochiai} and Jaccard
\cite{abreu2007accuracy}, they share the same basic insight, i.e.,
code elements mainly executed by failed tests are more suspicious.
%Although various SBFL techniques have been proposed, e.g., Tarantula \cite{jones2001visualization}, SBI \cite{liblit2005scalable}, Ochiai \cite{abreu2006evaluation} and Jaccard \cite{abreu2007accuracy}, they share the same basic insight, i.e., code elements mainly executed by failed tests are more suspicious.
%The input of SBFL is the coverage information of all tests and the output is a ranked list of code elements (e.g., statements or methods) according to their descending order of suspiciousness values calculated by specific formula.
The Mutation-based Fault Localization (MBFL), e.g.,~\cite{budd1981mutation,MUSE,musco2017large,zhang2010test,
  zhang2013injecting},
aims to additionally consider mutated code in fault
localization.
%since code elements covered by failed/passed tests may
%not have impact on the corresponding test outcomes.
The examples of MBFL are Metallaxis \cite{papadakis2012using,
  Metallaxis} and MUSE \cite{MUSE}.
%Mutation-based Fault Localization (MBFL), e.g.,~\cite{moon2014ask, zhang2013injecting,budd1981mutation, zhang2010test}, aims to additionally consider impact information for fault localization. Since code elements covered by failed/passed tests may not have any impact on the corresponding test outcomes, e.g., Metallaxis \cite{papadakis2012using, papadakis2015metallaxis} and MUSE \cite{moon2014ask}.
%
%typical MBFL techniques
%use mutation testing \cite{budd1981mutation, zhang2010test, musco2017large} to simulate the impact of each
%code element for more precise fault localization, e.g., Metallaxis \cite{papadakis2012using, papadakis2015metallaxis} and MUSE \cite{moon2014ask}.
%The first general MBFL technique, Metallaxis [24, 26] is based on the
%following intuition: if one mutant has impacts on failed tests (e.g.,
%the tests outcomes change after mutation), its corresponding code
%element may have caused the test failures
%
%The more recent MUSE [23] technique has similar insights: (1)
%mutating faulty elements may cause more failed tests to pass than
%mutating correct elements; (2) mutating correct elements may
%cause more passed tests to fail than mutating faulty elements.
%
Learning-to-Rank (LtR) has been used to improve fault
localization~\cite{b2016learning, TraPT, FLUCCS, MULTRIC}.
%The basic idea is to learn the potential faulty locations via combining suspiciousness values computed by various fault localization techniques (i.e., features).
MULTRIC \cite{MULTRIC} combines different suspiciousness values from SBFL. Some work combines SBFL suspiciousness values with
other information, e.g., program invariant \cite{b2016learning} and source code complexity information \cite{FLUCCS}, for more effective LtR in FL.
TraPT \cite{TraPT} combines suspiciousness values from both SBFL and MBFL.
%Various other machine learning techniques, and even
Neural networks have been applied to fault
localization~\cite{briand2007using,wong2009bp,zhang2017deep,zheng2016fault}. However,
they mainly work on the test coverage summarization scores, which has
clear limitations (e.g., it cannot distinguish elements covered by
both failing and passing test cases)~\cite{TraPT}, and are usually
studied on artificial faults or small programs. DeepFL~\cite{DeepFL}
was shown to improve the method-level FL approach TraPT~\cite{TraPT}.
%, by 36.54\% in terms of Recall at Top-1.
%Some other deep learning based fault localization has been applied
%•Neural attribution for semantic bug localization in student
%programs" (NeurIPS 2019).  Rishabh Singh, Benjamin Livshits, and
%Benjamin Zorn. Melford: Using Neural Networks to Find Spreadsheet
%Errors
%
{\tool} is also related to CNN-FL~\cite{cnn-fl}, which feeds the
original coverage matrix with passing/failing information into a CNN
model. CNN-FL is theoretically equivalent to {\em Base} model in
Table~\ref{fig:sensitivity_2}, without any matrix enhancements, test
cases ordering, statement dependencies, and code representations.

%, i.e.,DeepRL4FL without
%(Matrix_Enhancement,Test_Ordering,Statement_Dependency,Code_Representation). DeepRL4FL
%performs better than Base. We’ll add this comparison.

\textbf{Code Representation Learning (CRL).}  The recent success in
machine learning has lead to much interest in applying machine
learning, especially deep learning, to program analysis and software
engineering tasks, such as automated correction for syntax
errors~\cite{Bhatia-2016}, spreadsheet errors
detection~\cite{barowy2018excelint,singh2017melford}, fuzz
testing~\cite{Patra-2016}, program synthesis~\cite{Amodio-2017}, code
clones~\cite{White-2016,Smith-2009,Li-2017}, program
summarization~\cite{Allamanis-2016,Mou-2014}, code
similarity~\cite{Alon-2018,Zhao-2018}, probabilistic model for
code~\cite{Bielik-2016}, and path-based code representation, e.g.,
Code2Vec~\cite{Alon-2018} and Code2Seq~\cite{alon2018code2seq}.
%In the above PL and SE tasks,
All the approaches learn code representations using different program
properties.  However, none of the existing fault localization
techniques has performed direct code modeling and learning on code
coverage information of the test cases for the FL purpose as in
{\tool}.

\section{Conclusion}
We propose a deep learning based fault localization (FL) approach,
\tool, to improve existing FL approaches.  The key ideas include (1)
treating the FL problem as the image recognition; (2) enhancing code
coverage matrix by modeling the relations among statements and failing
test cases; (3) combining code coverage representation learning with
statement dependencies, and the code representation learning for usual
suspicious code.
%5fully exploiting all code coverage information by combining
%enhanced matrices with dynamically and statically generated statements
%dependency.
%Through empirical studies, \tool can localize the most bugs from a
%well known benchmark Defects4J compared with all studied baselines.
%It can also detect bugs in C projects with similarly high accuracy.
Our empirical evaluation shows that our model advances the
state-of-the-art baseline approaches.

%Our empirical studies show that DeepRL4FL outperforms the studied baseline and localizes 245 bugs from Defects4J, a well-known benchmark, which is the most in the FL literature. DeepRL4FL improves the top-1 results of baselines by up to 491.7\%. Furthermore, our sensitivity analysis shows that each of our designed components positively contributes to DeepRL4FL's high performance.

%Because our approach needs to do analysis on the relationship in order to locate the faults, very short method cannot provide enough information to make our approach work well.

\section*{Acknowledgment}
This work was supported in part by the US National Science
Foundation (NSF) grants CCF-1723215, CCF-1723432, TWC-1723198,
CCF-1518897, and CNS-1513263.

%\newpage

%\input{sections/conce}

%\input{sections/exploratory}
\balance

\bibliographystyle{IEEETrans}
%\citestyle{acmauthoryear} 
\bibliography{sections/FL}

%\begin{thebibliography}{00}
%\bibitem{b1} G. Eason, B. Noble, and I. N. Sneddon, ``On certain integrals of Lipschitz-Hankel type involving products of Bessel functions,'' Phil. Trans. Roy. Soc. London, vol. A247, pp. 529--551, April 1955.
%\bibitem{b2} J. Clerk Maxwell, A Treatise on Electricity and Magnetism, 3rd ed., vol. 2. Oxford: Clarendon, 1892, pp.68--73.
%\bibitem{b3} I. S. Jacobs and C. P. Bean, ``Fine particles, thin films and exchange anisotropy,'' in Magnetism, vol. III, G. T. Rado and H. Suhl, Eds. New York: Academic, 1963, pp. 271--350.
%\bibitem{b4} K. Elissa, ``Title of paper if known,'' unpublished.
%\bibitem{b5} R. Nicole, ``Title of paper with only first word capitalized,'' J. Name Stand. Abbrev., in press.
%\bibitem{b6} Y. Yorozu, M. Hirano, K. Oka, and Y. Tagawa, ``Electron spectroscopy studies on magneto-optical media and plastic substrate interface,'' IEEE Transl. J. Magn. Japan, vol. 2, pp. 740--741, August 1987 [Digests 9th Annual Conf. Magnetics Japan, p. 301, 1982].
%\bibitem{b7} M. Young, The Technical Writer's Handbook. Mill Valley, CA: University Science, 1989.
%\end{thebibliography}
%\vspace{12pt}
%\color{red}
%IEEE conference templates contain guidance text for composing and formatting conference papers. Please ensure that all template text is removed from your conference paper prior to submission to the conference. Failure to remove the template text from your paper may result in your paper not being published.

\end{document}